\documentclass[onecolumn,article]{IEEEtran}
\usepackage{times}

\usepackage[top=0.6in, bottom=0.7in, left=0.6in, right=0.6in]{geometry}
\usepackage{amsmath}  
\usepackage{amssymb}  
\usepackage{mathrsfs} 

\usepackage{theorem}  
\usepackage{cite}     
\usepackage{comment}  

\usepackage{upref}
\usepackage{amsfonts}

\usepackage{verbatim}

\usepackage{bbm}

\usepackage[dvipsnames,usenames]{color}
\usepackage{enumerate}

\usepackage{graphicx}
\usepackage{subfigure}

\usepackage{latexsym}

\usepackage[draft,ulem=normalem]{changes}
\usepackage[all]{xy}

\newcommand{\be}[1]{\begin{equation}\label{#1}}
\newcommand{\ee}{\end{equation}}

\newcommand{\bc}{\begin{center}}
\newcommand{\ec}{\end{center}}

\newcommand{\cB}{{\cal B}}
\newcommand{\cC}{{\cal C}}

\newcommand{\cG}{{\cal G}}
\newcommand{\cH}{{\cal H}}

\newcommand{\cP}{{\cal P}}

\newcommand{\cU}{{\cal U}}

\newcommand{\bfa}{{\boldsymbol a}}
\newcommand{\bfb}{{\boldsymbol b}}
\newcommand{\bfc}{{\boldsymbol c}}
\newcommand{\bfd}{{\boldsymbol d}}
\newcommand{\bfe}{{\boldsymbol e}}
\newcommand{\bff}{{\boldsymbol f}}

\newcommand{\bfh}{{\boldsymbol h}}

\newcommand{\bfs}{{\boldsymbol s}}

\newcommand{\bfu}{{\boldsymbol u}}
\newcommand{\bfv}{{\boldsymbol v}}
\newcommand{\bfw}{{\boldsymbol w}}
\newcommand{\bfx}{{\boldsymbol x}}
\newcommand{\bfy}{{\boldsymbol y}}
\newcommand{\bfz}{{\boldsymbol z}}
\newcommand{\bfA}{{\mathbf A}}

\newcommand{\bfM}{{\mathbf M}}

\renewcommand{\le}{\leqslant}
\renewcommand{\leq}{\leqslant}
\renewcommand{\ge}{\geqslant}
\renewcommand{\geq}{\geqslant}

\newcommand{\Z}{\mathbb{Z}}

\newcommand{\one}{{\mathbf 1}}

\newcommand{\Cref}[1]{Co\-rol\-la\-ry\,\ref{#1}}

\newcommand{\nchoosekv}[2]{\left((#1) \, ; \, (#2)\right)}
\newcommand{\nchoosek}[2]{\left(\begin{array}{c}#1\\#2\end{array}\right)}

\theoremstyle{plain} \theorembodyfont{\normalfont\slshape}

\newtheorem{thm}{Theorem$\!$}
\newenvironment{theorem}{\begin{thm}\hspace*{-1ex}{\bf.}}{\end{thm}}

\newtheorem{prop}[thm]{Proposition$\!$}
\newenvironment{proposition}{\begin{prop}\hspace*{-1ex}{\bf.}}{\end{prop}}

\newtheorem{lem}[thm]{Lemma$\!$}
\newenvironment{lemma}{\begin{lem}\hspace*{-1ex}{\bf.}}{\end{lem}}

\newtheorem{cor}[thm]{Corollary$\!$}
\newenvironment{corollary}{\begin{cor}\hspace*{-1ex}{\bf.}}{\end{cor}}

\newtheorem{prob}[thm]{Problem$\!$}

\newtheorem{defi}[thm]{Definition$\!$}

\newtheorem{claim}{Claim}

\theorembodyfont{\normalfont}

\newtheorem{exam}{Example$\!$}
\newenvironment{example}{\begin{exam}\hspace*{-1ex}{\bf .}}{\end{exam}}

\newtheorem{remrk}{Remark$\!$}

\definecolor{Codecolor}{named}{White}  

\newcommand{\Copen}{\mbox{\{\kern-5.50pt\{}}
\newcommand{\Cclose}{\mbox{\}\kern-5.50pt\}}}
\newcommand{\Cslash}{\mbox{$\backslash\kern-6.02pt\backslash$}}

\begin{document}
\title{Asymmetric Lee Distance Codes for DNA-Based Storage}
\author{
  \IEEEauthorblockN{
    Ryan~Gabrys\IEEEauthorrefmark{1}\IEEEauthorrefmark{4}, 
    Han~Mao~Kiah\IEEEauthorrefmark{3},~and
    Olgica~Milenkovic\IEEEauthorrefmark{4}}\\ \thanks{This work was supported in part by the NSF STC Class 2010 CCF 0939370 grant and the Strategic Research Initiative (SRI) Grant conferred by the University of Illinois, Urbana-Champaign. Parts of the results were presented at the Information Theory Workshop (ITW) in Jerusalem, 2015, and the International Symposium on Information Theory (ISIT), 2015.}
   \IEEEauthorblockA{
	 \IEEEauthorrefmark{4} Coordinated Science Laboratory, University of Illinois, Urbana-Champaign, USA\\
	\IEEEauthorrefmark{1}Spawar Systems Center San Diego, Code 532, USA\\
	 \IEEEauthorrefmark{3} School of Physical and Mathematical Sciences, Nanyang Technological University, Singapore\\
	 Emails: ryan.gabrys@navy.mil,  gabrys@illinois.edu, hmkiah@ntu.edu.sg,  milenkov@illinois.edu}  
	 \vspace{-10mm}}

\maketitle

\begin{abstract} We introduce a new family of codes, termed asymmetric Lee distance (ALD) codes, designed to correct errors arising in DNA-based storage systems and systems with parallel string transmission protocols. ALD codes are defined over a quaternary alphabet and analyzed in this particular setting, but the derived results hold for other alphabet sizes as well. 
Our technical contributions are two-fold: First, we derive upper bounds on the size of the codes under the ALD metric based on linear programming techniques. Second, we propose a number of code constructions which imply lower bounds.
\end{abstract}


\section{Introduction}\label{sec:aub}

Codes for classical symmetric channels with single-sequence inputs and single-sequence outputs have been extensively studied in the past and a number of diverse solutions such as algebraic codes~\cite{R06}, codes on graphs -- e.g., LDPC codes~\cite{RU01} -- and polar codes~\cite{TA13} have been proposed to address the underlying communication problem. Similar advances have been reported for parallel channels~\cite{HSST13}, under the rather common assumption that the individual channels introduce uncorrelated errors, as well as for channels with asymmetric error properties~\cite{K81}. In both scenarios, the alphabet size of the codes is restricted by the system design, and often, input sequences are de-interleaved or represented as arrays over smaller alphabets in order to enable more efficient transmission. Far less is known about channels that operate on several sequences at the same time and introduce correlated and asymmetric symbol errors. The goal of this work is to introduce and analyze one such scenario, motivated by an abstraction of emerging read channels used for DNA-based storage systems~\cite{K15,Y15a}. 

DNA-based storage systems have recently been implemented using biochemical methods for synthesizing (writing) the desired digital content in DNA~\cite{C12,G13,Y15a}, and reading the content via DNA sequencing systems. DNA synthesis may be performed with small error rates, the dominant error events being simple substitutions~\cite{Y15b}. The read process introduces errors that cannot be modeled using standard information-theoretic channels. For example, the high throughput Illumina sequencing platform may be viewed as a channel that has a single sequence as its input and a multiset of the input sequence substrings as its output~\cite{K15}. On the other hand, fast evolving, miniaturized nanopore sequencers such as MinION~\cite{M15} introduce context-dependent, bursty deletion errors. 

\begin{figure}[htb!]
\begin{center}
\includegraphics[scale=0.23]{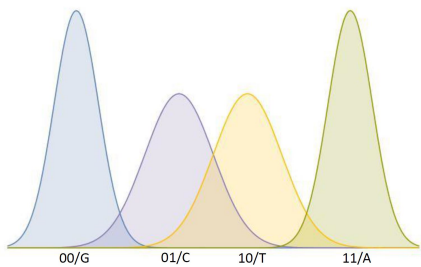} 
\end{center}
\vspace{-12mm}
\caption{Nanopore sequencers: detection system output distribution ($y$-axis) versus current drop in nA ($x$-axis), along with a mapping of binary words of length two to the DNA alphabet, as adopted in our analysis.}
\vspace{-2mm}\label{fig:DNAsymbols}
\end{figure}

The goal of this work is to propose designs for error-correcting codes capable of handling ``asymmetric'' errors arising in new solid state nanopore sequencing systems. In these systems, the DNA strings are read one symbol (nucleotide) at the time, and the identity of the symbol is determined by the drop of an otherwise fixed current caused by the nucleotide passing through the sequencing pore~\cite{radenovic15,K11}. Since nucleotides have different atomic structures, they lead to different current drop responses. A stylized depiction of one such system response to the four nucleotides {\tt A,T,G,C} is shown in Figure~\ref{fig:DNAsymbols}.

As may be seen from Figure~\ref{fig:DNAsymbols}, the impulse response distributions corresponding to {\tt T} and {\tt C} exhibit significant overlap, making the symbols highly confusable. In contrast, the impulse responses corresponding to {\tt A} and {\tt G} are widely separated and are very unlikely to be confused. The pairs of nucleotides {\tt G} and {\tt T} and {\tt A} and {\tt C} are only slightly confusable. Using a sequencer with the given symbol response would lead to \emph{asymmetric errors}, in so far that one would have high incidence rates of {\tt T-C} substitution errors, a significantly smaller incidence rate of {\tt G-T} and {\tt A-C} substitution errors, and a very low incidence rate of {\tt A-G} substitution errors.

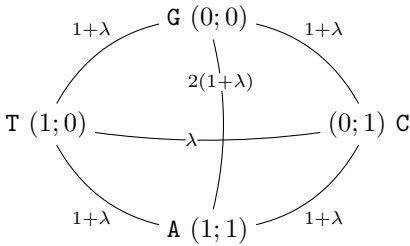
\begin{figure}
\centerline{
\xymatrix{
& {\tt G}\ (0;0) \ar@/^1pc/@{-}[dr]^{1+\lambda} \ar@/^/@{-}[dd]|(0.3){2(1+\lambda)}\\
{\tt T}\ (1;0)\ar@/^1pc/@{-}[ur]^{1+\lambda} \ar@/_/@{-}[rr]|(0.45){\lambda}&& (0;1)\ {\tt C}\ar@/^1pc/@{-}[dl]^{1+\lambda} \\
&{\tt A}\ (1;1)\ar@/^1pc/@{-}[ul]^{1+\lambda}\\
}
}
\caption{A weighted confusion graph for the code alphabet $\{{{\tt A,T,G,C}\}}$, involving $n=1$ symbols. The symbols are encoded as pairs of bits.}
\label{fig:metric}
\vspace{-0.2in}
\end{figure}

To abstract this error model, we introduce an asymmetric confusion graph shown in Figure~\ref{fig:metric}. 
There, we used the following binary encoding of the DNA alphabet: {\tt G} $\to (0;0)$, {\tt A} $\to (1;1)$, {\tt C} $\to (0;1)$ and {\tt T} $\to (1;0)$. The labels of the edges of the confusion graph depend on a positive integer parameter  $\lambda$ and are chosen to have a simple form that inversely correlates with the probability of the corresponding substitution error\footnote{Note that one can choose different functions of $\lambda$ for the edge labels, as long as the resulting distance $d$ satisfies $d((0;0),(1;1))<d((0;0),(1;0))<d((0;0),(1;1))$. The proposed functions $\lambda,\lambda+1,2(\lambda+1)$ were chosen for simplicity of analysis and based on the fact that for $\lambda=1$, they produce weights that are powers of two, namely $1,2,4$.}. As an example, the pairs of bits $(0;0)$ and $(1;0)$ are twice more likely to be confused than the pairs of bits $(0;0)$ and $(1;1)$. Hence, the distance ``induced'' by the former pair is half the distance ``induced'' by the latter pair. Clearly, one could adopt a significantly more general model for the labels of the confusion graph which more precisely captures the actual errors in the sequencing devise; but such a model would introduce unnecessary notational inconvenience and lead to tedious calculations that obscure the actual ideas behind the analysis.

\begin{figure}[htb!]
\begin{center}
\includegraphics[trim = 0mm 45mm 0mm 45mm, clip,scale=0.23]{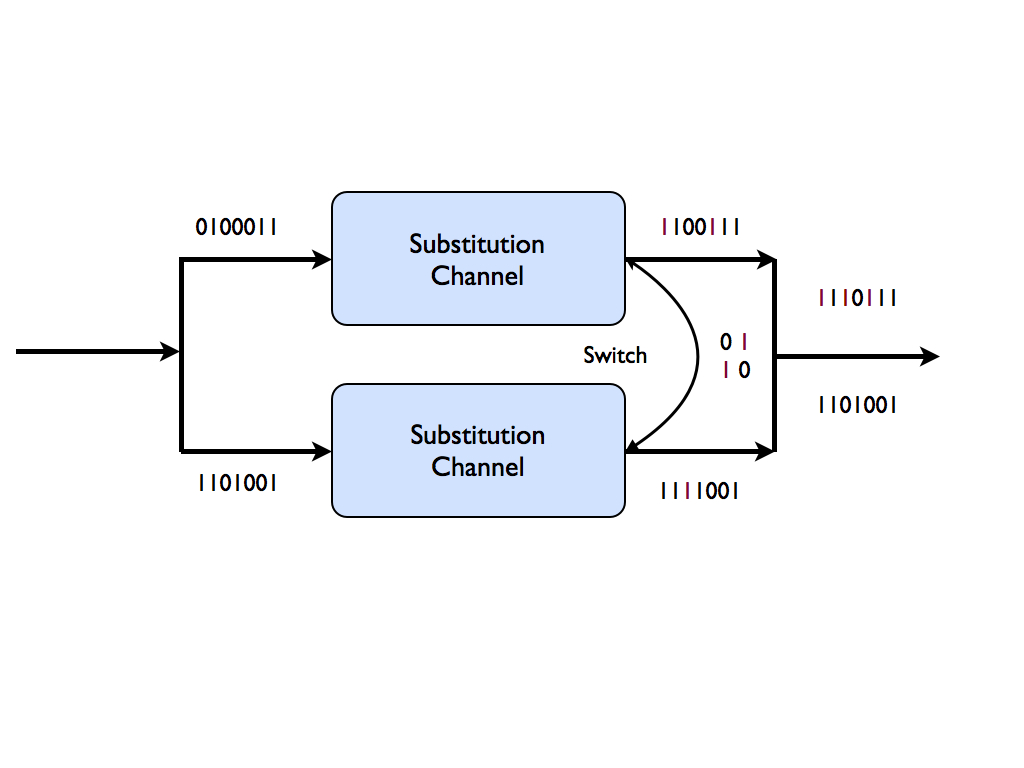} 
\end{center}
\vspace{-9mm}
\caption{A pair of channels with individual substitution errors, and outputs which may be switched. In the given example, the outputs of the channels at position three are switched.}
\vspace{-5mm}\label{fig:channel}
\end{figure}

The confusion graph in Figure~\ref{fig:metric} has another interpretation in terms of a parallel channel model, interesting in its own right. To motivate this interpretation, consider Figure~\ref{fig:channel} which depicts a transmission model where two binary input sequences are simultaneously passed through \emph{two channels} that introduce substitution errors. Simultaneous errors in both strings are less likely than individual string errors. In addition to the substitution errors, the outputs of the channels may be \emph{switched} -- in other words, the label of the channel from which the output symbol originated may be in error. The confusion graph for this type of channel is also captured by the model of Figure~\ref{fig:metric}, where the vertices are indexed by pairs of bits denoting the inputs into the two channels. Such switching errors may arise due to loss of synchronization in de-interleaving or due to the lack of channel side information.

The central problems of the presented work are to evaluate the ultimate coding limits under the distance induced by the symbol metric of Figure~\ref{fig:metric} and to describe combinatorial constructions for \emph{pairs} of binary sequences (codewords) that are at a sufficiently large ``asymmetric Lee distance'' (ALD) from each other. As discussed in the sections to follow, the ALD equals a scaled version of the Lee distance with an additional modification that two symbols are deemed to be of higher importance than the other two symbols. The two distinguished symbols capture the uncertainty about the actual ordering of the readouts in the parallel channel or the fact that two nucleotide impulse responses tend to show a high degree of overlap. The ALD was not previously studied in the literature, although its relationship with the Lee metric allows one to adapt certain proof methods~\cite{A82} used for the latter distance to the case of the ALD. Nevertheless, the asymmetry of the ALD also calls for developing new, specialized coding techniques that enable deriving tight bounds on the code size and constructing codes that approach these bounds. 

The contributions of this paper are three-fold. First, we introduce a new channel model and the corresponding distance measure that has applications in macromolecular storage. Second, we derive upper bounds on the size of codebooks under the newly introduced distance using recently proposed linear programming (LP) approaches on hypergraphs~\cite{KK13} and LP methods based on alphabet extensions and Delsarte-like techniques~\cite{delsarte}. Third, we describe a number of linear and nonlinear code constructions and accompanying decoding procedures. 

The paper is organized as follows. Section~\ref{sec:problem} contains a formal definition of the ALD. Section~\ref{sec:NAUB} describes how to use linear LP methods on hypergraphs in order to derive non-asymptotic upper bounds on codes in the ALD. Section~\ref{sec:UB} introduces a new LP framework that uses specialized alphabet embeddings. Sections~\ref{sec:constructions} and \ref{sec:nonconstructions} present new code constructions for the ALD.

\section{Preliminaries and Problem Formulation}\label{sec:problem}

For $a,b \in \mathbb{F}_2$, the indicator function $\mathbbm{1}(a,b)$ takes the value $\mathbbm{1}(a,b)=1$ if $a=b,$ and zero otherwise. Similarly, an $n$-variate indicator function for a collection of $n$ symbols $a_1,\ldots,a_n$ over $\mathbb{F}_2$ takes the value $\mathbbm{1}(a_1,\ldots,a_n)=1$ if $a_1=a_2=\ldots=a_n,$ and zero otherwise. As an illustration, $\mathbbm{1}(0,0,0)=1,$ while $\mathbbm{1}(0,1,1)=0$. 

Consider next four sequences $\bfa = (a_1, \ldots, a_n), \bfb = (b_1, \ldots, b_n), \bfc = (c_1, \ldots, c_n), \bfd = (d_1, \ldots, d_n) \in \mathbb{F}_2^n$, paired as $(\bfa;\bfb)$ and $(\bfc;\bfd)$. The pairing of two binary sequences of length $n$ results in one sequence over a quaternary alphabet as illustrated by the following example $((0,1,0);(1,1,1))=((0;1),(1;1),(0;1))$ where $((0;1),(1;1),(0;1)) \in (\mathbb{F}_2^2)^3$.

For a positive integer $\lambda$, the ALD $d_{\lambda}((\bfa;\bfb),(\bfc;\bfd))$ between two pairs of binary sequences $(\bfa;\bfb),(\bfc;\bfd)$ is defined as 
\begin{align}\label{eq:metric}
d_{\lambda}((\bfa;\bfb),(\bfc;\bfd))=&\sum_{i=1}^n \Big ( (1+\lambda)\left( \mathbbm{1}(a_i, b_i) + \mathbbm{1}(c_i, d_i) \right) + \\ \nonumber
&\lambda \mathbbm{1}(a_i, \bar{b}_i, \bar{c}_i, d_i)- 2(1+\lambda) \mathbbm{1}(a_i, b_i, c_i, d_i) \Big ), 
\end{align}
where $\bar{a}$ denotes the complement of the bit $a$.

The confusion graph for a code of length $n=1$ under the ALD matches the one depicted in Figure~\ref{fig:metric}: For example, it is easy to see that $d_{\lambda}((0;0),(1;1))$ equals
\begin{align}
(1&+\lambda)\left( \mathbbm{1}(0, 0) + \mathbbm{1}(1, 1)\right)+\lambda \mathbbm{1}(0, 1, 0, 1) -2(1+\lambda) \mathbbm{1}(0,0,1,1) \notag \\
&=(1+\lambda)(1+1)+\lambda\;0-2(1+\lambda)\;0=2(1+\lambda). \notag
\end{align}

It can be verified that $d_{\lambda}((\bfa;\bfb),(\bfc;\bfd))$ is a metric with respect to the pairs of binary sequences in the argument: $d_{\lambda}$ is non-negative and symmetric, and it satisfies the triangle inequality.  
From (\ref{eq:metric}) and the confusion graph, we also observe that the ALD is asymmetric in terms of the costs of different transitions, and in particular, complementary pairs are treated differently than non-complementary pairs. Furthermore, when pairs are complementary, the distance depends on the binary weight of the pairs. The choice of the parameter $\lambda$ governs the degree of asymmetry. 

In order to highlight the relationship between $d_{\lambda}((\bfa;\bfb),(\bfc;\bfd))$ and the Lee distance, let us first recall that the Lee distance between two symbols $a,b$ over $\mathbb{Z}_4$ is defined as $d_L(a,b)=\min(|a-b|,4-|a-b|)$. Next, let us introduce a mapping $\mathcal{Z}: \mathbb{F}_2^2 \to \mathbb{Z}_4$ such that $(0; 0) \to 1$, $(1;0) \to 0$, $(0;1) \to 2$ and $(1;1) \to 3$. 
By changing the weight between $(1;0)$ and $(0;1)$ from $\lambda$
to $2(1+\lambda)$, we arrive at a scaled version of the Lee distance $d_{L}(a,b)$ between two symbols $a,b \in \mathbb{Z}_4$, i.e.,
\begin{align}\label{eq:LeeinMap}
(1+\lambda) \, \cdot \, \min\{{4-|a-b|,|a-b|\}}=(1+\lambda) \, \cdot \, d_{L}(a,b).
\end{align}
More precisely, for two sequences $\bfz_1=(\bfa;\bfb),\bfz_2=(\bfc;\bfd)$ viewed over $\mathbb{Z}_4^n$, we have
\begin{align}\label{eq:LeeinMap}
&d_{\lambda}(\bfz_1,\bfz_2)=(1+\lambda) \cdot d_{L}(\bfz_1,\bfz_2) - \\
&(2+\lambda) \sum_{i:\,\{{z_{1,i},z_{2,i}\}} = \{{0,2\}}}\, (1-\mathbbm{1}(z_{1,i},z_{2,i})) \leq (1+\lambda) \cdot d_{L}(\bfz_1,\bfz_2). \nonumber
\end{align}

\begin{figure}
\centerline{
\xymatrix{
& 1 \ar@/^1pc/@{-}[dr]^{1+\lambda} \ar@/^/@{-}[dd]|(0.3){2(1+\lambda)}\\
0 \ar@/^1pc/@{-}[ur]^{1+\lambda} \ar@/_/@{-}[rr]|(0.40){2(1+\lambda)}&& 2 \ar@/^1pc/@{-}[dl]^{1+\lambda} \\
&3 \ar@/^1pc/@{-}[ul]^{1+\lambda}\\
}
}
\caption{A weighted confusion graph for the Lee distance scaled by $1+\lambda$.}
\label{fig:metric2}
\vspace{-0.2in}
\end{figure}
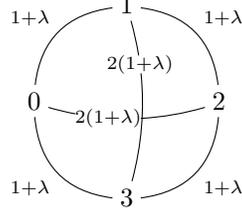

Simple inspection of the transition diagram for the ALD also reveals that 
$$d_{\lambda}(z_1,z_2) \geq \frac{\lambda}{2} d_{L}(z_1, z_2). $$

Furthermore, the Lee distance over $\mathbb{Z}_4$ equals the Hamming distance over $\mathbb{Z}_2^2$ under Gray mappings from
$\mathbb{F}_2^2$ and $\mathbb{Z}_4$. Hence, since $\mathcal{Z}$ is a Gray map, we have
$$ d_{\lambda}(z_1, z_2) \geq \frac{\lambda}{2} d_H(z_1, z_2). $$

Note that the distance between pairs of symbols $(a;b), (c;d) \in \mathbb{F}^2_2$ under the ALD depends on the exact values of the bits in $(a;b)$ and $(c;d)$ and not on their sum. For example, if $\lambda=1$, $(a;b) = (0;0)$ and $(c; d) = (1;1)$ then $d_{\lambda}((a;b),(c;d))=4$; on the other hand, for the same choice of the parameter $\lambda$ and for $(a; b) = (1;0)$ and $(c;d) =(0;1)$, we have $d_{\lambda}((a;b),(c;d))=1$, despite the fact that in both cases we have $(a; b)+(c; d)=(a+c;b+d)=(1;1)$. At the same time, for both choices of paired vectors, the underlying Lee distance equals two, as the Lee distance for the given mapping $\mathcal{Z}$ only depends on the Hamming weight of $(a;b) + (c;d)$. The connection between the ALD, Lee metric, and asymmetric error-correcting codes will be used in some of our subsequent derivations.

In what follows, we refer to an error that causes a transition between $(1;0)$ and $(0;1)$ as a \textit{Class 1 error}; similarly, we refer to an error that causes a single substitution in one of the input strings as a \textit{Class 2 error}. An error that causes a transition between $(0;0)$ and $(1;1)$ is referred to as a \textit{Class 3 error}. Note that based on Figure~\ref{fig:metric}, an edge in the confusion graph corresponding to a Class 1 error has weight $\lambda$, an edge corresponding to a Class 2 error has weight $1+\lambda$, while an edge corresponding to a Class 3 error has weight $2(1+\lambda)$. 

Let $d$ be a positive integer. We say that two pairs of sequences $(\bfa;\bfb), (\bfc;\bfd) \in \mathbb{F}^n_2 \times \mathbb{F}^n_2$ are \textit{(d,$\lambda$)-distinguishable} if their ALD $d_{\lambda}$ is at least $d$; conversely, we say that two pairs of sequences are \textit{(d, $\lambda$)-indistinguishable} if their ALD is less than $d$. Furthermore, for an ordered pair of sequences $({\bfa};{\bfb}) \in \mathbb{F}_2^n \times \mathbb{F}_2^n$, we let 
\begin{align}\label{eq:eball}
\cB_{(r, \lambda)}(\bfa;\bfb) := \{ (\bfc;\bfd) \in \mathbb{F}_2^n \times \mathbb{F}_2^n : d_{\lambda}( (\bfa;\bfb),(\bfc;\bfd)) \leq r \} 
\end{align}
denote the set of pairs of sequences that are $(r+1,\lambda)$-indistinguishable from $(\bfa;\bfb)$. 
We also use $A_\lambda(n, d)$ to denote the largest number of $(d, \lambda)$-distinguishable sequences of length $n$. 

\begin{example}\label{ex:intro} Let $\nchoosekv{0, 1, 1}{1, 1, 1} \in \mathbb{F}_2^3 \times \mathbb{F}_2^3.$ Then, the following eight vector pairings
\begin{align*}
& \nchoosekv{0, 1, 1}{1, 1, 1}, \nchoosekv{1,1,1}{1,1,1}, \nchoosekv{0,1,1}{0,1,1}, \\
&\nchoosekv{1,1,1}{0,1,1}, \nchoosekv{0,0,1}{1,1,1}, \nchoosekv{0,1,1}{1,0,1},\\
&\nchoosekv{0,1,0}{1,1,1}, \nchoosekv{0,1,1}{1,1,0}
\end{align*}
are $(3,1)$-indistinguishable from $\nchoosekv{0, 1, 1}{1, 1, 1}$. For example, the ALD between $\nchoosekv{1,1,1}{1,1,1}$, the second pair in the above list, and $\nchoosekv{0,1,1}{1,1,1}$, the chosen pair, equals $1+\lambda=2$, and this distance arises from the pair of symbols in the first positions, $(1;1)$ and $(0;1)$.

On the other hand, there are seven pairs of sequences $(3,1)$-indistinguishable from $\nchoosekv{0,0,0}{0,0,0}$, namely
\begin{align*}
& \nchoosekv{0, 0, 0}{0, 0, 0}, \nchoosekv{1,0,0}{0,0,0}, \nchoosekv{0,0,0}{1,0,0}, \\
&\nchoosekv{0,1,0}{0,0,0}, \nchoosekv{0,0,0}{0,1,0}, \nchoosekv{0,0,1}{0,0,0}, \\
&\nchoosekv{0,0,0}{0,0,1}.
\end{align*}
Note that the sum of the two vectors in the first example equals $\bfa+\bfb=(0, 1, 1)+(1, 1, 1)=(1,0,0)$, and hence has Hamming weight one. On the other hand, the sum of the two vectors in the second example equals $\bfa+\bfb=(0, 0, 0)+(0, 0, 0)=(0,0,0)$ and has Hamming weight zero. 
 \end{example}
Although the ALD depends on the exact values of its arguments ${\bfa},{\bfb}$, we show next that $|\cB_{(r, \lambda)}(\bfa;\bfb)|$ only depends on the Hamming weight of ${\bfa}+ {\bfb}$, where the sum is taken modulo $2$. This claim parallels a result for the $Z$-channel, which asserts that the size of the error sphere is a function of the Hamming weights of the transmitted sequences~\cite{K81}. To draw upon this parallel, we subsequently denote the Hamming weight of ${\bfa}+ {\bfb}$ by $w(\bfa;\bfb)$. Clearly, $w(\bfa; \bfb) = d_H(\bfa, \bfb)$, where $d_H(\bfa, \bfb)$ denotes the Hamming distance between $\bfa$ and $\bfb$.
\begin{lemma} Let $(\bfa;\bfb) \in \mathbb{F}_2^n \times \mathbb{F}_2^n$ be a pair of sequences with weight $w(\bfa; \bfb)$. Then, $|\cB_{(r,\lambda)}(\bfa;\bfb)|$ depends on $(\bfa;\bfb)$ only through $w(\bfa; \bfb)$.
\end{lemma}
\begin{IEEEproof} From (\ref{eq:metric}) and (\ref{eq:eball}), one can see that $|\cB_{(r,\lambda)}(\bfa; \bfb)|$ equals the number of paired sequences $(\bfc; \bfd) \in \mathbb{F}_2^n \times \mathbb{F}_2^n$ that satisfy
\begin{align*}
&\sum_{i=1}^n \Big( (1+\lambda)\left( \mathbbm{1}(a_i, b_i) + \mathbbm{1}(c_i, d_i) \right) + \\ \nonumber
&\lambda \mathbbm{1}(a_i, \bar{b}_i, \bar{c}_i, d_i)- 2(1+\lambda) \mathbbm{1}(a_i, b_i, c_i, d_i) \Big )\leq r.
\end{align*}
First, assume that $\bfa$ and $\bfb$ are fixed and let $E=\{i : a_i =b_i\}$ and $D=\{i : a_i \neq b_i\}$. 
We are interested in the number of paired sequences $(\bfc; \bfd) \in \mathbb{F}_2^n \times \mathbb{F}_2^n$ that satisfy 
\begin{align*}
&\sum_{i \in E} (1+\lambda)(1 + \mathbbm{1}(c_i, d_i)) - 2(1+\lambda) \mathbbm{1}(a_i, c_i, d_i) + \\
&\sum_{i \in D} (1+\lambda) \, \mathbbm{1}(c_i, d_i) +  \lambda \mathbbm{1}(a_i,\bar{c}_i,d_i) \leq r.
\end{align*}
The above inequality may be rewritten as
\begin{align*}
&(1+\lambda)\, \sum_{i=1}^{n} \mathbbm{1}(c_i, d_i) - 2(1+\lambda) \sum_{i \in E} \, \mathbbm{1}(a_i, c_i, d_i) + \\
&\lambda \sum_{i \in D} \mathbbm{1}(a_i,\bar{c}_i,d_i) \leq r -(1+\lambda) \, |E|.
\end{align*}
The first term in the above sum does not depend on the choice of $\bfa$. For $i \in E$, independent on the choice of $a_i$, the second sum provides a non-zero contribution if and only if $a_i=c_i=d_i$. For $i \in D$, independent on the choice of $a_i$, the third sum provides a non-zero contribution if and only if $a_i=\bar{c}_i=d_i$. Hence, it follows that the number of choices for $\bfc$ and $\bfd$ satisfying the inequality depends only on $|E|$ and $|D|$, and consequently on $w(\bfa;\bfb)$ only, as $w(\bfa;\bfb)=|D|$ and $|E|+|D|=n$.
\end{IEEEproof}

Next, we turn our attention to deriving an expression for $\mathcal{V}_{\lambda}(n, w, r): = |\cB_{(r, \lambda)}(\bfa;\bfb)|$, with $w(\bfa ; \bfb) = w$. 
Recall that $\cB_{(r, \lambda)}(\bfa;\bfb)$ was defined in (\ref{eq:eball}) and let 
$$ \mathbf{S}_{\lambda}(n,w,r) := \{ ({\bfc};{\bfd}) \in \mathbb{F}_2^n \times \mathbb{F}_2^n : d_{\lambda}( (\bfa;\bfb),(\bfc;\bfd)) = r \},$$
with $S_{\lambda}(n,w,r)=|\mathbf{S}_{\lambda}(n,w,r)|$.
Clearly, $\mathcal{V}_{\lambda}(n,w,r) = \sum_{j=0}^r S_{\lambda}(n,w,j).$ The next lemma establishes a formula for $S_{\lambda}(n,w,r)$ which may be subsequently used to determine $\mathcal{V}_{\lambda}(n,w,r)$. For simplicity of notation, we use $k \ast \ell \ast m \ast \lambda = r$ to denote $(2k+\ell)(1+\lambda) + \lambda m = r$, and similarly, $k \ast \ell \ast m \ast \lambda \leq r$ to denote $(2k+\ell)(1+\lambda) + \lambda m \leq r$.

\begin{lemma} For positive integers $n,w$ and $r$ such that $w \leq n$,
$$ S_{\lambda}(n,w,r) = \sum_{k,\ell,m: \, k \ast \ell \ast m \ast \lambda = r} \nchoosek{w}{m} \nchoosek{n-w}{k} \nchoosek{n-k-m}{\ell} 2^\ell. $$
\end{lemma}
\begin{IEEEproof} We count the number of sequences $(\bfc; \bfd) \in \mathbb{F}_2^n \times \mathbb{F}_2^n$ such that $d_{\lambda}( (\bfa; \bfb); (\bfc; \bfd) )  = r,$ where $w(\bfa; \bfb) = w$. Suppose that $(\bfc; \bfd)$ is the result of $m$ Class 1 errors, $k$ Class 3 errors, and $\ell$ Class 2 errors occurring in $(\bfa; \bfb)$. We first arbitrarily choose the locations of any of the Class 1 errors from among the $w$ locations in $(\bfa; \bfb)$ for which $a_1 \neq b_1$. There are $\nchoosek{w}{m}$ choices for the locations. Similarly, 
we can choose the locations of the Class 3 errors amongst the $n-w$ locations where $a_1 = b_1$. There are $\nchoosek{n-w}{k}$ choices for these locations. The locations of the Class 2 errors are chosen among the remaining $n-k-m$ locations in $(\bfa; \bfb)$, and this choice of locations is coupled with the choice of which of the $2$ available bits is changed (recall that a Class 2 error causes a substitution error in one of the two input strings at any given location). There are $\nchoosek{n-k-m}{\ell} 2^\ell$ choices for the Class 2 errors, which proves the statement in the lemma.
\end{IEEEproof}

\begin{corollary}  For positive integers $n,w$ and $r$ such that $w \leq n$, 
$$ \mathcal{V}_{\lambda}(n,w,r) = \sum_{k,\ell,m: \, k \ast \ell \ast m \ast \lambda \leq r} \nchoosek{w}{m} \nchoosek{n-w}{k} \nchoosek{n-k-m}{\ell} 2^\ell. $$
\end{corollary}

The next corollary will be useful for the derivations in the sections to follow.

\begin{corollary}\label{cor:balls} For positive integers $n,w$ and $r$ such that $w \leq n$,
it holds that $\mathcal{V}_{\lambda}(n,w,r) \geq \mathcal{V}_{\lambda}(n,w-1,r)$.  
\end{corollary}
\begin{IEEEproof} To prove the claimed result, consider the ratio 
\begin{align*}
\frac{\nchoosek{w}{m} \nchoosek{n-w}{k} \nchoosek{n-k-m}{\ell} 2^\ell }{ \nchoosek{w-1}{m} \nchoosek{n-w-1}{k} \nchoosek{n-k-m}{\ell} 2^\ell } = \frac{w(n-w)}{ (w-m) (n-w-k)}.
\end{align*}
Since $w \geq w-m$ and $n-w \geq n-w-k$ for all possible parameter choices, one has $\frac{w(n-w)}{ (w-m) (n-w-k)} \geq 1$. Given that
$$\mathcal{V}_{\lambda}(n,w,r) = \sum_{k \ast \ell \ast m \ast \lambda \leq r} \nchoosek{w}{m} \nchoosek{n-w}{k} \nchoosek{n-k-m}{\ell} 2^\ell$$
it clearly follows that $\frac{\mathcal{V}_{\lambda}(n,w,r)}{\mathcal{V}_{\lambda}(n,w-1,r)} \geq 1$. Thus, $\mathcal{V}_{\lambda}(n,w,r) \geq \mathcal{V}_{\lambda}(n,w-1,r),$ as desired.
\end{IEEEproof}

\section{Upper Bounds from Hypergraphs}\label{sec:NAUB}

In what follows, we focus our attention on non-asymptotic upper bounds derived based on a recent optimization on hypergraphs framework. We start our exposition by introducing the relevant notation borrowed from~\cite{KK13}.

\subsection{Hypergraph Preliminaries}

A hypergraph $\cH$ is an ordered pair of sets $(V,E)$, where $V$ denotes the set of vertices and where $E$ denotes the set of hyperedges, i.e., selected subsets of $V$. A \emph{matching} in $\cH$ is a set of disjoint hyperedges, and we use $\nu(\cH)$ to denote the size of the largest matching, i.e., the maximum number of disjoint hyperedges in $ \cH$. The number $\nu(\cH)$ is also known as the matching number of the hypergraph $\cH$. A \emph{fractional matching} of a hypergraph is a function $f$ with domain $E$ and range $[0,1]$ that assigns a value $f(e)$ to each hyperedge $e$ so that $\sum_{e \in E(v)}\, f(e)\leq 1$ for each vertex $v$ and its set of incident hyperedges $E(v)$. If we restrict the range of $f$ to $\{{0,1\}}$, the fractional matching reduces to the indicator function of a matching. The \emph{fractional matching number} of a hypergraph $\cH$, denoted by $\nu^*(\cH)$, equals
$$\nu^*(\cH)=\sup_{f: f \;\text{a fractional matching}} \; \sum_{e \in E}\, f(e),$$
or equivalently
$$\nu^*(\cH)=\sup\left\{\sum_{i=1}^{|E|} \, \textbf{z}_i \, : \textbf{M} \cdot \textbf{z} \leq 1, \textbf{z} \in\mathbb{R}_+^{|E|}\right\},$$
where $\textbf{M}$ denotes the vertex-hyperedge incidence matrix of the hypergraph.

A subset of the vertex set that has a nonempty intersection with every edge is called a \emph{transversal} of the hypergraph $\cH$, and the smallest size of a transversal is called the transversal number of the hypergraph, denoted by $\tau(\cH)$. Similarly as for the case of matchings, one may define a \emph{fractional transversal} as a function $g$ with domain $V$ and range $[0,1]$ that assigns a value $g(v)$ to each vertex $v$ so that $\sum_{v \in V(e)}\, g(v) \geq 1$ for each edge $e$ and its set of incident vertices $V(e)$. If we restrict the range of $g$ to $\{{0,1\}}$, the fractional transversal reduces to the indicator function of a transversal. The \emph{fractional transversal number} of a hypergraph $\cH$, denoted by $\tau^*(\cH)$, equals
$$\tau^*(\cH)=\inf_{g: g \; \text{a fractional transversal}} \; \sum_{v \in V}\, g(v),$$
or equivalently
$$\tau^*(\cH)=\inf\left\{\sum_{i=1}^{|V|} \, \textbf{z}_i \, : \textbf{M}^T \cdot \textbf{z} \geq 1, \textbf{z} \in\mathbb{R}_+^{|V|}\right\}.$$ 
Using duality arguments, it can be shown that $\nu^*(\cH)=\tau^*(\cH)$.

Next, fix two positive integers $r$ and $\lambda$ and recall the definition of $\cB_{(r,\lambda)}$ from (\ref{eq:eball}).
We introduce the following hypergraph
$$ \cH_{(r,\lambda), n} = (\mathbb{F}_2^n \times \mathbb{F}_2^n, \{ \cB_{(r,\lambda)}(\bfa;\bfb): (\bfa;\bfb) \in \mathbb{F}_2^n \times \mathbb{F}_2^n \}), $$
with vertex set $\mathbb{F}_2^n \times \mathbb{F}_2^n$ and edges that represent subsets of vertices within ALD balls of radius $r$ around some vertex. 
Then, the following result follows directly from~\cite{KK13}.
 
\begin{lemma} \label{lem:kk} Given the hypergraph $ \cH_{(r,\lambda), n}$, one has
$$A_{\lambda}(n,2r+1) = \nu(\cH_{(r,\lambda), n}) \leq \tau^{*}(\cH_{(r,\lambda), n}). $$
\end{lemma}
Intuitively, the lemma asserts the fact that a matching in the hypergraph corresponds to a packing of balls in the 
ALD metric, as no two hyperedges (balls) are allowed to intersect. Thus, the matching number is the largest 
number of disjoint balls that may be packed in the corresponding space, and hence it is equal to the largest size of an ALD code.

We find the following reformulation of the aforementioned bound useful~\cite{KK13}. Let $\bfA_{(r,\lambda)}$ be a square 
matrix with $4^n$ rows and columns, indexed by the elements of $\mathbb{F}_2^n\times \mathbb{F}_2^n$. 
We set $\bfA_{(r,\lambda)}(i,j)=1$ if 
$(\bfx_{i,1};\bfx_{i,2})\in \cB_{(r,\lambda)}(\bfx_{j,1};\bfx_{j,2}),$ and zero otherwise, for any two pairs 
$(\bfx_{i,1};\bfx_{i,2}), (\bfx_{j,1};\bfx_{j,2}) \in \mathbb{F}_2^n\times \mathbb{F}_2^n$. Observe that $\bfA_{(r,\lambda)}$ is the adjacency matrix of the ALD $r$-confusion graph.
 
Let $\mathbb{R}_{+}$ denote the set of non-negative reals. Then,
{\small
\begin{align}\label{eq:transversal}
\tau^{*}(\cH_{(r,\lambda), n}) = \min \Bigg\{ \sum_{i=1}^{4^n} w_i : {\bfw} \in \mathbb{R}_{+}^{4^n}, \, \bfA^T_{(r,\lambda)} \cdot {\bfw} \geq {\bf1} \Bigg\}.
\end{align}
}
For simplicity, we subsequently write  $\tau^{*}\left(\bfA_{(r,\lambda)}\right)$ to refer to $\tau^{*}(\cH_{(r,\lambda), n})$, as $\tau^{*}(\cH_{(r,\lambda), n})$ is a function of $\bfA_{(r,\lambda)}$ only. Note that \eqref{eq:transversal} is an LP involving $4^n$ constraints, which may be prohibitively large to solve even for moderate values of $n$. We show next how to significantly reduce this number of constraints by using \emph{hypergraph symmetries}~\cite{FVY14}.

Following the approach of~\cite{FVY14}, define a directed graph $\cG_{(r,\lambda)}$ on the vertex set $\mathbb{F}_2^n\times\mathbb{F}_2^n$, with an arc from $(\bfx_1;\bfx_2)$ to $(\bfy_1;\bfy_2)$ iff $(\bfy_1;\bfy_2) \in\cB_{(r,\lambda)}(\bfx_1;\bfx_2)$. An automorphism $\pi$ of $\cG_{(r,\lambda)}$ is a permutation in $\mathbb{S}_{4^n}$ (the symmetric group of order $(4^n)!$) that operates on the vertex set and preserves adjacencies. The set of all automorphisms of $\cG_{(r,\lambda)}$ is denoted by $ {\rm Aut}(\cG_{(r,\lambda)})$ and represents a subgroup of
$\mathbb{S}_{4^n}$. The group ${\rm Aut}(\cG_{(r,\lambda)})$ induces a relation $R$ on the vertices of $\cG_{(r,\lambda)}$ in the following manner: $(x,y) \in R$ iff there exists a $\pi \in {\rm Aut}(\cG)$ such that $\pi(x) = y.$ Clearly, $R$ is an equivalence relation. Hence, given a subgroup $H$ of ${\rm Aut}(\cG_{(r,\lambda)})$ of order $n_H$, we may use $H$ to partition the vertex set into $n_H$ equivalence classes
$\{V_{H,1}, \ldots, V_{H,n_H}\}$.

With a change in the indices of the corresponding expression in~\cite{FVY14}, let $\bfA_{H,(r,\lambda)}$ be an $n_H \times n_H$ matrix indexed by the elements of the subgroup $H$, with entries $\bfA_{H,(r,\lambda)}(i, j)$ defined according to
\[
\frac{\Big |\Big \{ \big (  (\bfx_1;\bfx_2),(\bfy_1;\bfy_2) \big) : (\bfx_1;\bfx_2) \in V_{H,j},\ (\bfy_1;\bfy_2) \in V_{H,i} \cap \cB_{(r,\lambda)}(\bfx_1;\bfx_2) \Big \} \Big |}{|V_{H,j}|},\, 1\le  i, j \le  n_H.
\]

We interpret the matrix $\bfA_{H,(r,\lambda)}$ as follows: One is given $n_H$ equivalence classes, representing the indices of the rows and columns of the matrix. The entry in row $i$ and column $j$ is the normalized number of pairs $(\bfx_1; \bfx_2),(\bfy_1; \bfy_2)$ such that $(\bfx_1; \bfx_2)$ is in the equivalence class $j$, and $(\bfy_1; \bfy_2)$ is in a subset of the equivalence class $i$ that comprises points in an $r$-ball around $(\bfx_1; \bfx_2)$. Normalization is performed with respect to the size of the $j$-th equivalence class, $|V_{H,j}|$.

Using the ALD hypergraph symmetries, we can simplify the statement in Lemma~\ref{lem:kk} and arrive at the following theorem which follows from the methods introduced in~\cite{FVY14,KK13}.
\begin{theorem} \label{th:kkbound}
Let $H$ be a subgroup of ${\rm Aut}(\cG_{(r,\lambda)})$ and let $\bfA_{H,(r,\lambda)}$ be as defined before. 
Then, 
$$ \tau^{*}(\bfA_{(r,\lambda)}) = \min \Bigg\{ \sum_{i=1}^{n_H} | V_{H,i} | \cdot w_i : \bfw \in \mathbb{R}_{+}^{n_H}, \bfA_{H,(r,\lambda)}^T \cdot {\bfw} \geq 1\Bigg\},$$
\noindent and $ A_{\lambda}(n,2r+1) \leq \tau^{*}(\bfA_{(r,\lambda)}).$
\end{theorem}


We next identify a set of automorphisms of $\mathcal{G}_{(r, \lambda)}$ that allow us to apply Theorem~\ref{th:kkbound} to the 
case of ALD codes. For every permutation $\sigma=(\sigma(1),\ldots,\sigma(n)) \in \mathbb{S}_n$ and for every $\bfz \in \mathbb{F}_2^n$, let 
\begin{align}\label{eq:pimap}
\pi_{\sigma, {\bfz}} : \mathbb{F}_2^n \times \mathbb{F}_2^n \to \mathbb{F}_2^n\times \mathbb{F}_2^n 
\end{align}
be a mapping such that for all $(\bfa;\bfb) \in \mathbb{F}_2^n \times \mathbb{F}_2^n$, $(\pi_{\sigma, {\bfz}}({\bfa};{\bfb}))_i = (a_{\sigma(i)}; b_{\sigma(i)})$ if $z_i = 0$ and $(\pi_{\sigma, \bfz}({\bfa};{\bfb}))_i = (\bar{a}_{\sigma(i)};\bar{b}_{\sigma(i)})$ otherwise. 
It is straightforward to see that $H= \{ \pi_{\sigma, \bfz} : \sigma \in \mathbb{S}_n, \bfz \in \mathbb{F}_2^n \}$ is a subgroup of ${\rm Aut}(\cG_{(r, \lambda)})$:
Coordinate permutations do not change the code, and complementation preserves the ALD symmetries of the pairings. 

The subgroup $H$ partitions the space $\mathbb{F}_2^n \times \mathbb{F}_2^n$ into $n+1$ equivalence classes, 
with $(\bfa; \bfb),(\bfc; \bfd) \in \mathbb{F}_2^n$ belonging to the same equivalence class if and only if $w(\bfa;\bfb)=w(\bfc;\bfd)$. 
The equivalence class containing a pair of vectors $\bfa$ and $\bfb$ of weight $w(\bfa;\bfb)=i$ has $2^n\, \binom{n}{i}$ elements.
Consequently,
{
\begin{align}
\tau^{*}(\bfA_{(r,\lambda)}) &= \min \Bigg\{ 2^n \cdot \sum_{\ell=0}^n \nchoosek{n}{\ell} \cdot w_\ell :  
 \label{eq:constraint} \\
 &\hspace{20mm}  {\bfw} \in \mathbb{R}_{+}^{n+1},\bfA_{H,(r,\lambda)}^T \cdot \bfw \geq 1 \Bigg\}. \notag
\end{align}
}

\begin{table}

\centerline{
\begin{tabular}{c|c c c c c c c}
$n\backslash d_{\lambda=1}$ &
3 & 5 & 7 & 9 & 11 & 13 \\
\hline
1 & 3 & 1 & 1 & 1 & 1 & 1 \\
2 & 9 & 2 & 2 & 1 & 1 & 1 \\
3 & 30 & 7 & 5 & 2 & 2 & 1 \\
4 & 99 & 21 & 13 & 5 & 4 & 2 \\
5 & 336 & 67 & 35 & 12 & 9 & 4 \\
6 & 1161 & 219 & 101 & 32 & 21 & 9 \\
7 & 4080 & 731 & 296 & 90 & 51 & 20 \\
8 & 14535 & 2483 & 895 & 258 & 130 & 50 \\
9 & 52377 & 8563 & 2783 & 771 & 348 & 127 \\
10 & 190557 & 29901 & 8890 & 2361 & 966 & 340 \\
\end{tabular}}
{\caption{Bounds on ALD code sizes based on the LP in (\ref{eq:constraint})} \label{table:program2}}
\end{table}

By using the automorphism group, 
the dimension of the weight vector $\bfw$ reduces from $4^n$, as given in~\eqref{eq:transversal}, 
to $n+1$, as given in~\eqref{eq:constraint}. Solutions of the LP of (\ref{eq:constraint}), for $n \leq 9$ and odd $d_{\lambda=1} \leq 13$ are listed in Table~I. 

To complete our analysis, we exhibit two feasible weight vectors for the optimization problem in~\eqref{eq:constraint} which lead to appropriate upper bounds on the size of ALD codes. 
\vspace{-0.07in}

\begin{theorem}\label{th:naivew} Let $\mathcal{V}_{\lambda}(n,w, r) = \mathcal{V}_{\lambda}(n, 0, r)$ for $w < 0$.
Suppose that $n,r,\lambda$ are positive integers, and set $\mu = \lfloor \frac{r}{1 + \lambda} \rfloor$. Let the elements of ${\bfw}=(w_0, w_1, \ldots, w_n) \in \mathbb{R}_+^{n+1}$ be given as
$$w_i = \frac{1}{\mathcal{V}_{\lambda}(n,i - \mu, r)}.$$
Then, Theorem~\ref{th:kkbound} leads to
\begin{equation*}
A_{\lambda}(n,2r+1) 
\leq  2^n \cdot \sum_{i=0}^n \nchoosek{n}{i} \cdot  \frac{1}{\mathcal{V}_{\lambda}(n,i - \mu, r)}.
\end{equation*}
\end{theorem}

\begin{IEEEproof} The result follows directly from Theorem~\ref{th:kkbound} and (\ref{eq:constraint}), provided that we can show that $\bfA_{H,(r,\lambda)}^T \cdot \bfw \geq 1$. For the given weight assignment, and for an arbitrary choice of an equivalence class  $V_{H,j}$ and $(\bfa; \bfb) \in V_{H,j}$, we have

\begin{align}
\sum_{\{((\bfa;\bfb),(\bfc;\bfd)) : (\bfa;\bfb) \in V_{H,j}, (\bfc;\bfd) \in V_{H,i} \cap \cB_{(r,\lambda)}(\bfa;\bfb) \}} \frac{w_{w(\bfc;\bfd)}}{|V_{H,j}|}  &= \sum_{(\bfc; \bfd) \in \cB_{\substack{(r,\lambda)}}(\bfa; \bfb)} \bfw_{w(\bfc; \bfd)} \nonumber \\
&=\sum_{(\bfc;\bfd) \in \cB_{(r, \lambda)}({\bfa}; {\bfb})} \frac{1}{\mathcal{V}_{\lambda}(n,w({\bfc};{\bfd}) - \mu, r)} \nonumber \\
&\geq \sum_{(\bfc;\bfd) \in \cB_{(r, \lambda)}({\bfa}; {\bfb})} \frac{1}{\mathcal{V}_{\lambda}(n,w({\bfa};{\bfb}), r)} \nonumber \\
&= \frac{\mathcal{V}_{\lambda}(n, w({\bfa};{\bfb}), r)}{\mathcal{V}_{\lambda}(n,w({\bfa};{\bfb}), r)}= 1, \nonumber
\end{align}
where the inequality follows from Corollary~\ref{cor:balls} and the observation 
that the weight of any element in $\cB_{(r,\lambda)}(\bfa; \bfb)$ is at most $w(\bfa; \bfb) + \mu$.
\end{IEEEproof}


Note that the choice of weights $\bfw$ from the previous theorem produces the best possible upper bound on $A_{\lambda=r}(n,2r+1)$ achievable via Theorem~\ref{th:kkbound}. This can be seen by noting that for $r=\lambda$, one may write
\begin{align*}
\tau^{*}(\bfA_{(\lambda, \lambda)}) =& \min \Bigg\{ 2^n \cdot \sum_{\ell=0}^n \nchoosek{n}{\ell} \cdot w_\ell : \\
  &\ \ \ \ w_\ell \cdot (\ell + 1) \geq 1, 0 \leq \ell \leq n, w_\ell \geq 0 \Bigg\}. 
\end{align*}
The bounds to follow are concerned with special choices of the problem parameters and do not involve solving LPs.
 
\begin{prop}\label{prop:optimal1}
Suppose that $r=\lambda$. Then 
\[\tau^*(\bfA_{(r, \lambda)})=2^n \cdot \sum_{\ell=0}^n \frac{\binom{n}{\ell}}{\ell+1}=\frac{2^n(2^{n+1}-1)}{n+1}.\]
Hence, $A_\lambda(n,2\lambda+1)\le {2^n(2^{n+1}-1)}/{(n+1)}$.
\end{prop}
\begin{IEEEproof} Note that when $r=\lambda$, $A_{H,(r, \lambda)}$ is a diagonal matrix with entries $a_{\ell, \ell} = \ell+1$ for $\ell \in \{0,1,\ldots, n\}$. Let $\bfw$ be a vector with entries $\bfw_{\ell}=(\frac{1}{\ell+1})$, $\ell=0,\ldots,n$. Then, $\bfA_{H,(r,\lambda)}^T \cdot \bfw = \one$ and $\bfw$ is a feasible solution of the underlying LP.
On the other hand, since $\bfA_{H,(r,\lambda)}$ is symmetric, $\bfA_{H,(r,\lambda)} \cdot \bfw = \one$ and 
$\bfw$ is a feasible solution of the dual LP. This proves the claimed result.
\end{IEEEproof}

The following proposition may be proved along the same lines as Proposition~\ref{prop:optimal1}. It relies on using sphere-packing arguments and a subgraph of the ALD confusion graph.

\begin{prop}\label{prop:simple}
For positive integers $r,\lambda$,
\[A_\lambda(n,2r+1)\le 2^n \cdot \sum_{\ell=0}^n \frac{\binom{n}{\ell}}{ \sum_{j=0}^{\lfloor r/\lambda \rfloor} \nchoosek{\ell}{j}}.\]
\end{prop}
\begin{IEEEproof} The proof follows from the simple observation that the vector $\bfw$ with entries
$$\bfw_{\ell}= \Bigg ( \frac{1}{\sum_{j=0}^{\lfloor r/\lambda \rfloor} \nchoosek{\ell}{j} } \Bigg ),\, \ell=0,\ldots,n,$$ 
satisfies $\bfA_{H,(r,\lambda)}^T \cdot \bfw \geq \one$, and is hence a feasible solution of the linear program.
\end{IEEEproof}

Consider next the special case when $\lambda =1$, and $r \geq 2$. To ease the notation, we introduce a function $K(n,r, i, j)$, such that
\begin{align*}
K(n,r, i, j)=&\sum_{\ell=0}^{\lfloor \frac{r}{2} \rfloor} \nchoosek{i}{\ell} \cdot 2^\ell \cdot \nchoosek{n-i}{j-i + \ell} \cdot 2^{j-i+\ell} \cdot \\
&\ \ \  \sum_{4k+m \leq r-2(j-i+2\ell)} \nchoosek{n-j-\ell}{k} \nchoosek{i-\ell}{m},
\end{align*}
for $j \geq i$, and
\begin{align*}
K(n,r, i, j)=&\sum_{\ell=0}^{ \lfloor \frac{r}{2} \rfloor} \nchoosek{n-i}{\ell} \cdot 2^\ell \cdot \nchoosek{i}{i-j+\ell} \cdot 2^{i-j+\ell} \cdot \\
&\ \ \ \ \sum_{4k+m \leq r-2(i-j+2\ell) } \nchoosek{j-\ell}{m} \nchoosek{n-i-\ell}{k},
\end{align*}
for $j < i$. We also assume that $K(n,r,i,j)$ $= 0$ whenever $r-2|i-j| < 0$. Given this setup, we may write {$\bfA^T_{H,(r, 1)}= \left( a_{i,j} \right)_{i=1,j=1}^{n+1}$}, where $a_{i,j} =K(n, r, i-1, j-1)$. The above claim follows from the observation that
$$K(n, r, i, j) = \Big| \Big\{ (\bfy_1; \bfy_2) : w(\bfy_1; \bfy_2) = j, (\bfy_1; \bfy_2) \in \cB_{(r,1)}(\bfx_1; \bfx_2) \Big \} \Big|,$$ 
where $w(\bfx_1; \bfx_2)=i$. To arrive at the expression for the above set-cardinality for the case $j \geq i$, one has to observe that one may choose $\ell$ locations for Class 2 errors from among the $i$  locations where $\bfx_1$ and $\bfx_2$ differ in value; one may then choose the locations of the remaining $j - (i -\ell)$ Class 2 errors from among the $n-i$ locations where $\bfx_1$ and $\bfx_2$ are equal in value. From the remaining locations, we may choose the locations of $k$ Class 3 errors from among positions in which the vectors agree in value, and $m$ Class 1 errors from among positions in which the vectors disagree in value. Similar reasonings apply for the case that $j < i$.


Next, we produce a weight assignment for (\ref{eq:constraint})
 by considering another matrix $\hat{\bfA}(n,r)$ related to $\bfA^T_{H,(r,1)}$ that takes advantage of the structure of $\bfA^T_{H,(r, 1)}$. The basic idea is to change some values of $\bfA^T_{H,(r,1)}$ so as to ensure that the diagonal elements of $\hat{\bfA}(n,r)$ are sufficiently large in order to invoke certain results regarding positive solutions of linear equations. The weights are then given as $\bfw = \hat{\bfA}(n,r)^{-1} \cdot \one$, and Theorem~\ref{th:weights1} establishes that $\bfw = \hat{\bfA}(n,r)^{-1} \cdot {\bf1}$ is a feasible weight assignment for \eqref{eq:constraint}.

The matrix $\hat{\bfA}(n,r)$ is constructed as follows: let $\hat{\bfA}(n,r)=\left( \hat{a}_{i,j} \right)_{i=1,j=1}^{n+1}$. Then,
$$\hat{a}_{i,j} = \min \Big (K(n, r, i-1, j-1), \frac{1}{r} (K(n,r, i-1, i-1) - 1) \Big )$$ 
for $j \neq i$, and $\hat{a}_{i,i}=K(n,r,i-1,i-1)$.
We find the following lemma useful for the proof of Theorem~\ref{th:weights1}.

\begin{lemma}\label{lem:positive} (c.f. \cite{K85}) Suppose that the matrix $\bfM=(m_{ij}) \in \mathbb{R}_{+}^{n \times n}$ contains no zero elements on the diagonal, and let ${\bfb} > 0$ be chosen arbitrarily. If for all $i \in \{1,\ldots, n\}$,
$$ b_i > \sum_{j=1, j \neq i}^n m_{ij} \frac{b_j}{m_{jj}}$$
then $\bfM$ is invertible and $\bfM^{-1} \bfb \geq 0$.
\end{lemma}

\begin{theorem}\label{th:weights1} For any two positive integers $n,r$, we have
$$ A_{\lambda=1}(n,2r+1) \leq 2^n \cdot \sum_{\ell=0}^n \nchoosek{n}{\ell} \cdot w_i, $$
where $\bfw = (w_0, \ldots, w_{n}) = \hat{\bfA}(n,r)^{-1} \cdot (1,1,\ldots,1)$.
\end{theorem}
\begin{IEEEproof} 
First, note that by the definition of the matrix $\hat{\bfA}(n,r)$, for all $i \in \{1, \ldots, n+1 \}$, one has

\begin{align*}
\sum_{j=1, j \neq i}^{n+1} \frac{\hat{a}_{i,j}}{\hat{a}_{j,j}} \leq \sum_{j=i - \lfloor \frac{r}{2} \rfloor, j \neq i}^{i + \lfloor \frac{r}{2} \rfloor} \frac{ K(n,r, i-1, i-1)-1}{ r  K(n,r, i-1, i-1)} < 1.
\end{align*}

Based on Lemma~\ref{lem:positive}, there exists a $\bfw$ with non-negative entries such that $\hat{\bfA}(n,r) \cdot \bfw = 1$. By definition, we also have $\bfA_{H,(r,1)}^T \geq \hat{\bfA}(n,r)$ and since $\hat{\bfw} \geq 0$, it follows that
$$ \bfA_{H,(r,1)}^T \cdot \bfw \geq \hat{\bfA}(n,r) \cdot \bfw = 1, $$
as desired. Thus, since $\bfA_{H,(r,1)}^T \cdot \bfw \geq 1$ and $\bfw \geq 0$, $\bfw$ is a feasible solution for the program in (\ref{eq:constraint}). 
\end{IEEEproof}

The results of our bounds on $A_{\lambda=1}(n,5)$, when $n \leq 15$, are listed in Table~\ref{table:UBHypergraphs}. In general, the bound from Proposition~\ref{prop:simple} provides the weakest upper bound on the code size while the bound from Theorem~\ref{th:weights1} 
best approximates the optimal weights solving the LP (\ref{eq:constraint}). Notice that the bound from Theorem~\ref{th:naivew} is weaker than the bound from Theorem~\ref{th:weights1} for all code lengths considered.

\begin{table}
\centerline{
\begin{tabular}{c|cccc}                                                    		   
Length	     & Bound from (\ref{eq:constraint}) & Theorem~\ref{th:naivew} & Proposition~\ref{prop:simple} 	& Theorem~\ref{th:weights1}  \\
\hline
5          &   67         &       427			      & 	254                  		& 112            \\
6          &    219      &       1079			&	793                  		& 354            \\
7          &    731       &      2750			&	2508              	      & 1170          \\ 
8	    &    2483     &       7181			&	8048                		& 3793          \\
9          &    8563     &     19485   		      &	26190              		& 12008        \\
10        &    29901   &     55529			&	86393                	      &  40400       \\
11        &    105490  &    166902			&	288649                	& 139753       \\
12        &   375448  &    527725			&	975954               	& 486642     \\
13        &   1346201 &   1742275       	 	&	3336118			      & 1665472     \\
14        &   4858171  &  5949948     		&	11518362		      & 5679816    \\
15        &  17631726 &  20833123		      & 	40130869	            & 19999983     \\
\end{tabular}}}
\vspace{0.07in}
{\caption{Comparison of Upper Bounds for $A_{\lambda=1}(n,5)$ \label{table:UBHypergraphs}}
\end{table}

In the next section, we consider another approach for deriving upper bounds on $A_{\lambda}(n,d).$ The method we use is a combination of novel symbol mapping techniques and the classical Delsarte-like LP approach. As will be shown by computer simulation, Delsarte-like bounds tend to outperform hypergraph methods for ALD codes for a number of parameter settings.

%

\section{Upper Bounds from Delsarte-Like Inequalities}\label{sec:UB}

We focus next on classical LP techniques for bounding the number of $(d,\lambda)$-distinguishable sequences $A_{\lambda}(n,d)$. The gist of our approach is to map the symbols from $\mathbb{F}_2 \times \mathbb{F}_2$ into symbols from some sufficiently large alphabet for which the distance constraints may be captured in a unique manner. We then proceed by introducing the complete distance enumerator for ALD codes, and use these enumerators to define an appropriate LP.

In order to explain why we need to map the symbols of $\mathbb{F}_2 \times \mathbb{F}_2$ into symbols from a larger alphabet, consider the following natural mapping onto $\mathbb{Z}_4$:
\begin{align*}
(0;0) &\mapsto 0, &
(0;1) & \mapsto 1, &
(1;0) &\mapsto 2, &
(1;1) & \mapsto 3.
\end{align*}
Since $2-1 \equiv 0-3 \bmod 4$, one cannot distinguish between the binary pairs $(1;0), (0;1)$ and $(0;0), (1;1)$ based on the difference of their images. To insure distinguishability of the differences of the images, 
one may consider using a mapping from $\mathbb{F}_2 \times \mathbb{F}_2$ into $\mathbb{F}_2 \times \mathbb{F}_2 \times \mathbb{F}_2$. Unfortunately, this mapping results in poor bounds, due to the fact that many of the LP constraints derived from this mapping are redundant. 

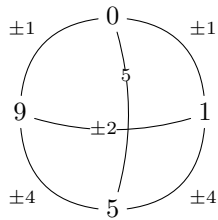
\begin{figure}[!h]
\centerline{
\xymatrix{
& 0\ar@/^1pc/@{-}[dr]^{\pm 1} \ar@/^/@{-}[dd]|(0.3){5}\\
9\ar@/^1pc/@{-}[ur]^{\pm 1} \ar@/_/@{-}[rr]|(0.45){\pm 2}&& 
1\ar@/^1pc/@{-}[dl]^{\pm 4} \\
&5 \ar@/^1pc/@{-}[ul]^{\pm 4}
}
}
\caption{The map $\phi^{(10)}$. The labels of the edges equal the difference of their defining vertices, taken modulo ten and reflected to lie in the interval $[-5,4]$.}\label{fig:metricten}
\end{figure}

Consider another mapping $\phi^{(10)}:\mathbb{F}_2\times\mathbb{F}_2\to \mathbb{Z}_{10}$ given by 
\begin{align*}
(0;0) &\mapsto 0, &
(0;1) & \mapsto 1, &
(1;0) &\mapsto 9, &
(1;1) & \mapsto 5.
\end{align*}
This mapping and the pairwise differences of the images are depicted in Figure~\ref{fig:metricten}. For any positive integer $n>1$, we extend the definition of the mapping $\phi^{(10)}:\mathbb{F}_2^n\times\mathbb{F}_2^n\to \mathbb{Z}_{10}^n$ in a natural manner as $\phi^{(10)}(\bfa;\bfb)=(\phi^{(10				)}(a_i;b_i))_{i=1}^n$. 

Let $\bfx \in \Z^n_{10}$. Define the profile of $\bfx$, denoted by 
$P(\bfx)$, as $(m_{x})_{x \in\Z_{10}}$, where
$m_{x}=\left|\{i: x_i=x\}\right|$. For example, the profile of $\bfx=(5,1,4,1) \in \mathbb{Z}_{10}^4$ equals 
$P(\bfx)=(0,2,0,0,1,1,0,0,0,0)$. With each $x \in\Z_{10}$, associate an indeterminate $z_{x}$. 
Given a collection of words $\cC\subseteq \mathbb{F}_2^n\times \mathbb{F}_2^n$, define the \textit{complete distance enumerator} of $\cC$ as
\begin{align*}
W_{\cC}(&z_{0},z_{1},\ldots,z_{9}) = \\ 
&\sum_{m_0,\ldots,m_9} w[m_{0},m_{1},\ldots,m_{9}]z_{0}^{m_{0}}z_{1}^{m_{1}}\cdots z_{9}^{m_{9}},
\end{align*}
where 
\begin{align*}
&w[m_{0},m_{1},\ldots,m_{9}]= \, \frac1{|\cC|} \big|\{(\bfa;\bfb),(\bfc;\bfd)\in\cC : \\
&P(\phi^{(10)}(\bfa;\bfb) - \phi^{(10)}(\bfc;\bfd))=(m_{0},m_{1},\ldots,m_{9})\}\big|,
\end{align*}
and where the differences are applied componentwise and evaluated modulo ten. 
\begin{claim} Let $\cC\subseteq \mathbb{F}_2^n\times \mathbb{F}_2^n$ with $d_\lambda(\cC)\ge d$ have a complete distance enumerator $\sum_{m_0,\ldots,m_9} w[m_{0},$ $\ldots,m_{9}]z_{0}^{m_{0}}$ $\cdots z_{9}^{m_{9}}$. Then, the following claims are true:
\begin{enumerate}[(i)]
\item $|\cC|=\sum_{m_0,\ldots,m_9} w[m_{0},\ldots,m_{9}]$, which states that the number of codewords may be retrieved by setting all variables in the complete weight enumerator to one;
\item $w[n,0,\ldots,0]=1$, which follows from the definition of the complete weight enumerator;
\item $w[m_{0},\ldots,m_{9}]=0$ whenever $m_{3}>0$ or $m_{7}>0$, which ensures that the edge labels $\pm 3$ are not allowed;
\item $w[m_{0},\ldots,m_{9}]=0$ if 
$(1+\lambda)(m_{1}+m_{4}+m_{6}+m_{9})+\lambda (m_{2} + m_8)+2(1+\lambda)m_{5}<d$, which captures the minimum distance constraint.
\end{enumerate}
\end{claim}

Let 
$$\chi({\bfx},{\bfy})= \exp(-2 \pi \iota \, (\bfx^T \cdot \bfy) / 10),$$ 
where $\iota$ denotes the imaginary unit, $\bfx,\bfy \in \mathbb{Z}_{10}^n$, and $\bfx^T \cdot \bfy = \sum_{i=1}^n x_i y_i$. Clearly, $\chi(\bfx,\bfy)=\prod_{i=1}^n\chi(x_i,y_i)$. For $\bfx \in \mathbb{Z}_{10}^n$, suppose that $P(\bfx)=(m_{0}, \ldots, m_{9})$ and define $\bfz(\bfx)=\prod_{i} z_{i}^{m_{i}}$.

\begin{lemma}\label{lem:delswitch}
Fix $\bfx \in \mathbb{Z}_{10}^n$ and suppose that $P({\bfx}) = (m_{0}, \ldots, m_{9})$. Then
\[\sum_{\bfy \in \mathbb{Z}^n_{10}} \bfz(\bfy)\chi(\bfx,\bfy)=F_{\textbf{z}}(m_{0},m_{1},\ldots,m_{9}),\]
where 
\begin{equation}
F_{\textbf{z}}(m_{0},m_{1},\ldots,m_{9}) = \prod_{j=0}^9 \left( \sum_{i=0}^9 z_i \chi(i,j) \right)^{m_j}.
\end{equation}
\end{lemma}

\begin{IEEEproof}
The claim follows by induction. The base case $n=1$ can be easily verified, while for $n \geq 2$, it suffices to observe that 
\small{\begin{align*}
\sum_{(y_1, y_2)} \bfz(y_1, y_2) &\chi \Big ( (x_1, x_2), (y_1, y_2) \Big)= \\
& \Big(\sum_{y_1} \bfz(y_1)\chi( x_1, y_1) \Big ) \cdot \Big (\sum_{y_2} \bfz(y_2)\chi(x_2,y_2) \Big ).
\end{align*}}
\end{IEEEproof}
\begin{theorem}\label{th:delsarte}
Let $\cC\subseteq \mathbb{F}_2^n\times \mathbb{F}_2^n$ and as before write 
$$\sum_{m_0, \ldots, m_9} w[m_{0},\, \ldots,m_{9}]z_{0}^{m_{0}}\cdots z_{9}^{m_{9}}$$ 
for the complete distance enumerator of $\cC$. Define $b[m_{0},\ldots,m_{9}]$ via
\begin{align*}
\sum_{m_0,\ldots,m_9} &w[m_{0},\ldots,m_{9}]F_{\textbf{z}}(m_{0}, \ldots, m_{9}) = \\ 
&\sum_{m_0,\ldots,m_9} b[m_{0},\ldots,m_{9}] z_{0}^{m_{0}} \cdots z_{9}^{m_{9}}.
\end{align*}
Then $b[m_{0},\ldots,m_{9}] \geq 0$, for all $m_{0},\ldots,m_{9}$.
\end{theorem}
\begin{IEEEproof}
For ease of notation, let $\bfx,\bfy \in \cC$ denote two codewords $(\bfx_1;\bfx_2)$ and $(\bfy_1;\bfy_2)$, respectively. 
Consider the following expression:
\[\sum_{\bfx,\bfy \in\cC}\sum_{\bfu \in \mathbb{Z}_{10}^n}\bfz(\bfu)\chi(\bfu,\phi^{(10)}(\bfx) - \phi^{(10)}(\bfy)).\]
On the one hand, using Lemma~\ref{lem:delswitch}, we may rewrite the given expression as 
\[ |\cC|\sum_{m_0, \ldots, m_9} w[m_{0},\ldots,m_{9}]F_{\textbf{z}}(m_{0},\ldots,m_{9}).\]
On the other hand, by switching the order of summation, one arrives at
\begin{align*}
&\sum_{\bfu \in \mathbb{Z}^n_{10}}\bfz(\bfu)\sum_{\bfx,\bfy\in\cC}\chi(\bfu,\phi^{(10)}(\bfx) - \phi^{(10)}(\bfy))\\
&=\sum_{\bfu \in \mathbb{Z}^n_{10}}\bfz(\bfu)\sum_{\bfx,\bfy\in\cC}\chi(\bfu,\phi^{(10)}(\bfx))\chi(\bfu,-\phi^{(10)}(\bfy))\\
&=\sum_{\bfu \in \mathbb{Z}^n_{10}} \bfz(\bfu)\left(\sum_{\bfx \in\cC}\chi(\bfu,\phi^{(10)}(\bfx))\right)\left(\sum_{\bfy\in\cC}\chi(\bfu,-\phi^{(10)}(\bfy))\right)\\
&=\sum_{\bfu \in \mathbb{Z}^n_{10}}\bfz(\bfu) \Big| \sum_{\bfx \in\cC}\chi(\bfu,\phi^{(10)}(\bfx)) \Big|^2.
\end{align*}
The proof follows since $\Big | \sum_{\bfx\in\cC}\chi(\bfu,\phi^{(10)}(\bfx)) \Big |^2 \, \geq 0$.
\end{IEEEproof}

As a consequence of the above results, an upper bound for $|\cC|$, with $d_\lambda(\cC)$ given, may be obtained from the following LP:
\begin{align*}
\text{maximize}  &\sum_{m_0,\ldots, m_9} w[m_{0},\ldots,m_{9}], \text{subject to\ } \\
		& w[n,0,\ldots,0]=1,\\
                 & w[m_{0},\ldots,m_{9}]=0, \text{ if } m_{3}>0 \text{ or } m_{7} > 0,\\
                 &w[m_{0},\ldots,m_{9}]=0, \text{ if } (1+\lambda)(m_{1}+ m_{4} \\
                 &\hspace{21.0ex}+m_{6}+m_{9}) + \lambda( m_{2} + m_8)\\
                 &\hspace{21.0ex}+2(1+\lambda)m_{5}<d,\\
                 & \sum_{m_0,\ldots,m_9} w[m_{0},\ldots,m_{9}]F_{\textbf{z}}(m_{0},\ldots,m_{9}) \ge 0.
\end{align*}
The multivariate polynomial inequality is to be interpreted as a nonnegativity constraint on the coefficients of the polynomials.
\begin{table*}
\centerline{
\begin{tabular}{c|cccc cccc cccc cccc cccc}
$n\backslash d_{\lambda=1}$ &
1 & 2 & 3 & 4 & 5 & 6 & 7 & 8 & 9 & 10 &
11 & 12 & 13 & 14 & 15 & 16 & 17 & 18 & 19 & 20 \\
\hline
1 & -- & -- & 2 & 2 \\
2 & -- & -- & -- & -- & 2 & 2 & 2 & 2 \\
3 & -- & -- & -- & -- & -- & -- & 4 & 4 & 2 & 2 & 2 & 2 \\
4 & -- & -- & -- & -- & -- & -- & -- & -- & 4 & 4 & 3 & 3 & 2 & 2 & 2 & 2 \\
5 & -- & -- & -- & -- & -- & -- & -- & -- & -- & -- & 6 & 6 & 3 & 3 & 2 & 2 & 2 & 2 & 2 & 2 \\
\end{tabular}
}
\vspace{0.06in}
\caption{Results of the Delsarte LP approach for $\lambda=1$}\label{tab:delsarte1}
\vspace{-0.05in}
\centerline{
\begin{tabular}{c|cccc cccc cccc cccc cccc}
$n\backslash d_{\lambda=1}$ &
1 & 2 & 3 & 4 & 5 & 6 & 7 & 8 & 9 & 10 &
11 & 12 & 13 & 14 & 15 & 16 & 17 & 18 & 19 & 20 \\
\hline
1 & -- & -- & 3 & 3 \\
2 & -- & -- & -- & -- & 2 & 2 & 2 & 2 \\
3 & -- & -- & -- & -- & -- & -- & 5 & 5 & 2 & 2 & 2 & 2 \\
4 & -- & -- & -- & -- & -- & -- & -- & -- & 5 & 5 & 4 & 4 & 2 & 2 & 2 & 2 \\
5 & -- & -- & -- & -- & -- & -- & -- & -- & -- & -- & 9 & 9 & 4 & 4 & 4 & 4 & 2 & 2 & 2 & 2 \\
\end{tabular}\label{table:program2}
}
\vspace{0.06in}
\caption{Results of (\ref{eq:constraint}) for $\lambda=1$}\label{tab:delsarte2}
\vspace{-0.12in}
\end{table*}


As can be seen from Tables~\ref{tab:delsarte1} and~\ref{tab:delsarte2}, for a number of cases where $d \geq 2n+1$, the bounds obtained from Delsarte's method are tighter than the bounds obtained via the LP techniques of Section~\ref{sec:NAUB}. For the case $d < 2n+1$, the output of the LP from Theorem~\ref{th:delsarte} was unbounded.

\section{Lower Bounds from Linear Code Constructions}\label{sec:constructions}

Throughout the remainder of the paper we present \emph{constructive} lower bounds on the sizes of codes under the ALD. As a starting point, we focus mainly on the case where $\lambda=1$ although many of the constructions can be extended to other parameter settings. Also, for simplicity, we first address the case where the minimum ALD is equal to three. 

For a positive integer $v$, let $H_{\bf{0,1}} \in \mathbb{F}_2^{v \times (2^v-2)}$ be a matrix which has as its columns all non-zero vectors from $\mathbb{F}_2^v,$ except for the all-ones vector. Write $H_{\bf{0,1}} = ({\bfh}_1, {\bfh}_2, \ldots, {\bfh}_{2^v-2}),$ where for $i \in \{1, \ldots, 2^{v}-2\}$, ${\bfh}_i$ represents the $i$-th column of $H_{\bf{0,1}}$. Furthermore, let ${\bf1}_v \in \mathbb{F}_2^{v \times 1}$ be the all-ones vector. 

In this setting, define $\cC_l(2^v -2)  \subseteq \mathbb{F}_2^{2^v-2} \times \mathbb{F}_2^{2^v-2}$ as
\begin{align*}
\hspace{-2.5ex}\cC_l(2^v-2) :=& \Big \{ ( {\bfa};{\bfb} )\in \mathbb{F}_2^{2^v-2} \times \mathbb{F}_2^{2^v-2} :  \\
& \sum_{i=1}^{2^v-2} a_i \cdot {\bfh}_i +  \sum_{i=1}^{2^v - 2} b_i \cdot {\bf1}_v = \bf0 \},
\end{align*}
where $({\bfa}; {\bfb} ) = ( (a_1, \ldots, a_{2^v-2}); (b_1, \ldots, b_{2^v-2}))$.

The subscript ``$l$'' in the definition of the code $\cC_l(2^v-2)$ serves to indicate that the code is a linear code. 
From the introductory discussion, observe that a code with $\lambda=1$ and minimum ALD distance three can either: 
\begin{enumerate}
\item Correct a single Class 1 error, or 
\item Detect a single Class 2 error. 
\end{enumerate}
In the following lemma, we show that the above defined code $\cC_l(2^v-2)$ can perform either 1) or 2). 

\begin{lemma}\label{lem:construct1} For any positive integer $n$, $d_{\lambda=1}(\cC_l(2^v-2)) \geq 3$. \end{lemma}
\begin{IEEEproof} Let $n = 2^v-2$. Suppose that $(\bfa;\bfb) \in \cC_l(2^v-2)$ was transmitted and that the pair of words $(\bfc;\bfd) \in \mathbb{F}_2^{n} \times \mathbb{F}_2^{n}$ was received, where $(\bfc;\bfd)$ is the result of at most one Class 2 error occurring in $(\bfa;\bfb)$. For the remainder of this proof, let $\bfs = \sum_{i=1}^{n} c_i \cdot {\bfh}_i + \sum_{i=1}^{n} d_i \cdot \one_v$ denote the decoding ``syndrome''. 
If a single Class 2 error occurred in position $j \in \{1,\ldots, n\}$ of $\bfa$, then  $\bfs = {\bfh}_j \ne \bf0$.
Otherwise, if a Class 2 error occurred in position $j$ of $\bfb$, then  $\bfs = \one_v \ne \bf0 $.
Clearly, if no Class 2 errors occurred, then $\bfs = {\bf0}$. Thus, $\cC_l(2^v-2)$ can detect whether a single Class 2 error has occurred by checking if $\bfs$ is non-zero.


Suppose that $(\bfa;\bfb) \in \cC_l(2^v-2)$ was transmitted and that the vector $(\bfc;\bfd) \in \mathbb{F}_2^{n} \times \mathbb{F}_2^{n}$ was received, where $(\bfc;\bfd)=( (c_1;d_1), (c_2;d_2), \ldots, (c_n;d_n))$ is the result of at most one Class 1 error in $(\bfa;\bfb)$. We describe next how to recover $(\bfa;\bfb)$ from $(\bfc;\bfd)$. Let $\tilde{\bfs} = \bfs + {\bf1}_v$, where $\bfs$ denotes the previously defined syndrome. 
Note that if $(\bfc;\bfd)$ is the result of a single Class 1 error in position $j$ of $(\bfa;\bfb)$, then 
$$\hspace{-2.0ex}\tilde{\bfs} =  \left( \sum_{i=1}^{n} c_i \cdot {\bfh}_i +  \sum_{i=1}^{n} d_i\cdot\one_v \right) + {\bf1}_v = {\bfh}_j + {\bf1}_v + {\bf1}_v = {\bfh}_j.$$
Otherwise, if no error occurred, $\tilde{\bfs} = \bfs + {\bf1}_v = {\bf1}_v$. From the above discussion, and by recalling that  ${\bfh}_j \neq {\bf1}_v$ by construction, it is clear that a decoder for $\cC_l(2^v-2)$ can recover $(\bfa;\bfb)$ from $(\bfc;\bfd)$, and $(\bfc;\bfd)$ from $\tilde{\bfs}$ as follows. If $\tilde{\bfs}={\bf1}_v$, then the decoder concludes that no errors have occurred. Otherwise, if $\tilde{\bfs} = {\bfh}_j$ for some $j \in \{1,\ldots, n \}$, then the decoder corrects a Class 1 error at position $j$.
\end{IEEEproof}

For a $\bfu \in \mathbb{F}_2^v$, let $\cC_{l,\bfu}(2^v -2)  \subseteq \mathbb{F}_2^{2^v-2} \times \mathbb{F}_2^{2^v-2}$ be a linear coset of $\cC_{l}(2^v -2)$ so that

\begin{align}\label{eq:cosetcode}
\hspace{-2.5ex}\cC_{l,\bfu}(2^v-2) :=& \Big \{ ( {\bfa};{\bfb} )\in \mathbb{F}_2^{2^v-2} \times \mathbb{F}_2^{2^v-2} : \notag \\
& \sum_{i=1}^{2^v-2} a_i \cdot {\bfh}_i +  \sum_{i=1}^{2^v - 2} b_i \cdot {\bf1}_v = \bfu \}.
\end{align}

The next result follows directly from Lemma~\ref{lem:construct1}. It will be used in the following section.

\begin{corollary}\label{cor:construct1lc} For any positive integer $n$ and $\bfu \in \mathbb{F}_2^v$, $d_{\lambda=1}(\cC_{l,\bfu}(2^v-2)) \geq 3$.
\end{corollary}

As a consequence of Lemma~\ref{lem:construct1} and Proposition~\ref{prop:optimal1}, we have 
$$ \small{\frac{4^n}{n+2} \leq A_{\lambda=1}(n,3) \leq \frac{2^n (2^{n+1} - 1)}{n+1}.} $$

We now turn our attention to the problem of constructing codes with $\lambda=1$ and minimum ALD equal to $d$, where $d>3$ is arbitrary. We first describe the code construction, and then proceed to provide a proof for its correctness. 

Let $H_d \in \mathbb{F}_2^{s \times 2n}$ be a parity check matrix for a code $\cC$ with Hamming distance $d$. Write $H_d = ({\bfh}_1' ,\ldots, {\bfh}_{2n}')$, where ${\bfh}_i'$ denotes the $i$-th column of $H_d$, and define $\cC_L(n) \subseteq \mathbb{F}_2^n \times \mathbb{F}_2^n$ according to

\begin{align*}
\hspace{-3.5ex} \cC_L(n) :=& \Big \{ ( {\bfa};{\bfb} )=( (a_1;b_1), \ldots, (a_n;b_n)) \in \mathbb{F}_2^{n} \times \mathbb{F}_2^{n} :  
\\
& \sum_{i=1}^{n} a_i \cdot {\bfh}_i' + \sum_{i=1}^{n} b_i \cdot ({\bfh}_i' + {\bfh}_{n+i}')   = \bf0 \Big \}.
\end{align*}

Similar to before, the subscript ``$L$'' on $\cC_L(n)$ refers to the fact that this is a linear code, and in order to distinguish it from the distance-three code $\cC_{l}$, we used the capital letter $L$.

We have the following result.

\begin{lemma}\label{lem:d3linear} For positive integers $n$ and $d \geq 4$, $d_{\lambda=1}(\cC_L(n)) \geq d$. \end{lemma}
\begin{IEEEproof}  Let $(\bfa;\bfb) \in \cC_L(n)$ for $d \geq 4$. Since $(\bfa;\bfb) \in \cC_L(n)$,  $\sum_{i=1}^{n} a_i \cdot {\bfh}_i' + \sum_{i=1}^{n} b_i \cdot ({\bfh}_i' + {\bfh}_{n+i}')   = \bf0 $. Therefore, if $(\bfc;\bfd) \in \cC_L(n)$, then $\sum_{i=1}^{n} c_i \cdot {\bfh}_i' + \sum_{i=1}^{n} d_i \cdot ({\bfh}_i' + {\bfh}_{n+i}')   = \bf0 $ as well. By combining these two equations, we deduce that given $(\bfa; \bfb) \in \cC_L(n)$, $(\bfc;\bfd) \in \cC_L(n)$ if and only if
\begin{align}\label{eq:lemd3linear1}
\sum_{i=1}^{n} (a_i+c_i) \cdot {\bfh}_i' + \sum_{i=1}^{n} (b_i+d_i) \cdot ({\bfh}_i' + {\bfh}_{n+i}')   = \bf0.
\end{align}

It remains to show that for any $(\bfc;\bfd) \in \mathbb{F}_2^n \times \mathbb{F}_2^n$ where  $d_{\lambda=1}( ({\bfa};{\bfb}); ({\bfc};{\bfd})) < d$, equation (\ref{eq:lemd3linear1}) does not hold. Suppose, in particular, that $(\bfc;\bfd)$ is the result of $\epsilon_1$ Class 1 errors, $\epsilon_2$ Class 2 errors, and $\epsilon_3$ Class 3 errors occurring in $(\bfa;\bfb)$, where $\epsilon_1 + 2 \epsilon_2 + 4 \epsilon_3 < d$. 
Then, 
$$\sum_{i=1}^{n} (a_i+c_i) \cdot {\bfh}_i' + \sum_{i=1}^{n} (b_i+d_i) \cdot ({\bfh}_i' + {\bfh}_{n+i}')$$ 
can be written as a sum of at most $\epsilon_1 + 2 \epsilon_2 + \epsilon_3$ columns of $H_d$. Since $\epsilon_1 + 2 \epsilon_2 + 4 \epsilon_3 < d$, and consequently $\epsilon_1 + 2 \epsilon_2 + \epsilon_3<d$, and since $H_d$ is a parity-check matrix of a code with Hamming distance $d$, we conclude that (\ref{eq:lemd3linear1}) does not hold. Therefore, $\sum_{i=1}^{n} c_i \cdot {\bfh}_i' + \sum_{i=1}^{n} d_i \cdot ({\bfh}_i' + {\bfh}_{n+i}')   \neq \bf0 $ and so $(\bfc;\bfd) \not \in \cC_L(n)$.
\end{IEEEproof}

We now comment on the size of the code $\cC_L(n)$ for the case when $d_{\lambda=1}=5$. More precisely, we compare the ALD codes $\cC_L(n)$ with distance five to codes constructed for the Hamming distance. In particular, we focus on (a) binary codes of length $2n$; and (b) the binary image of quaternary codes of length $n$. 

For (a), since a code with minimum ALD equal to five must be able to correct four errors (resulting from two Class 1 errors), we require a binary code $\cC$ that has minimum Hamming distance at least nine, so that one has
$$|\cC| \leq \frac{2^{2n}}{\sum_{j=0}^4 \nchoosek{2n}{j}},$$ 
which follows from the sphere packing bound. For $2n=2^v-2$, where $v \geq 2$ is a positive integer, 
$$|\cC_L(n)| \geq \frac{4^{n}}{(2n+2)^2},$$ 
provided that one-bit shortened primitive binary BCH codes of minimum Hamming distance $5$ are used as defining codes $H_d$. 
In this case, $|\cC| \leq |\cC_L(n)|$. Proposition~\ref{prop:simple} for the case $\lambda=1$ and distance five, and the lower bound on $|\cC_L(n)|$ imply that for $2n=2^v-2$, with $v>2$ an integer, it also holds that
$$ \frac{4^{n}}{(2n+2)^2} \leq A_{\lambda=1}(n,5) \leq \frac{3 \cdot 2^n}{(n+1)(n+2)}\left( 2^{n+2} - n - 3 \right).$$

Next, consider the second case, (b). A quaternary code $\cC$ with minimum ALD distance five has to have Hamming distance at least five. Writing out the sphere packing bound  for quaternary codes with Hamming distance five, we arrive at
$$|\cC|\le  \frac{4^n}{ \sum_{j=0}^2 \binom{n}{j} 3^j}.$$
For large enough $n$, this value is strictly smaller than $ \frac{4^n}{(2n+2)^2}$, which is a lower bound on $|\cC_L(n)|$ for $d_{\lambda=1}=5$. 
More precisely,
$$ \small{|\cC| \leq \frac{4^n}{ \sum_{j=0}^2 \nchoosek{n}{j} 3^j} < \frac{4^n}{(2n+2)^2} \leq |\cC_L(n)|,}$$
whenever $v \geq 5$. Hence, the previously proposed constructions also outperform the ``direct approach'' with a quaternary alphabet.

\section{Lower Bounds from Non-Linear Code Constructions}\label{sec:nonconstructions}

In what follows, we present constructions of non-linear codes under the ALD. In many instances, the new code constructions result in codebooks larger than their linear counterparts. We first consider the simplest case when $d=2$ and $\lambda=1,$ and codes that can \emph{detect} one single Class 1 error. Then, we extend our ideas to the case where the minimum ALD of the code is equal to three and $\lambda=1$. For this parameter case, we improve upon the construction in Lemma~\ref{lem:d3linear} whenever the block length is $n=2^v - 2$, for any positive integer $v \geq 5$. General code constructions are given at the end of the section.

Recall that the weight of a vector pairing $(\bfa;\bfb)=((a_1, \ldots, a_n);(b_1, \ldots, b_n)) \in \mathbb{F}_2^n \times \mathbb{F}_2^n$ is defined as $w(\bfa;\bfb) = | \{ i : a_i \neq b_i \}|$. Let $\mathcal{S}(\bfa;\bfb) = (a_{j_1}, \ldots, a_{j_{w({\bfa}, {\bfb})}})$ be a subsequence of $\bfa$ obtained by deleting all the entries of $\bfa$ on which $\bfa$ and $\bfb$ agree. In other words, if $\mathcal{S}(\bfa;\bfb) = (a_{j_1}, \ldots, a_{j_{w({\bfa}, {\bfb})}})$, then ${j_1} < {j_2} < \cdots < j_{w( {\bfa}, {\bfb} )}$ is the largest set of indices for which $a_{j_1} \neq b_{j_1}, a_{j_2} \neq b_{j_2}, \ldots, a_{j_{w( {\bfa}, {\bfb} )}} \neq b_{j_{w( {\bfa}, {\bfb} )}}$. Furthermore, we denote the set of indices supported in $\mathcal{S}(\bfa;\bfb)$, i.e., the set of indices for which $\bfa$ and $\bfb$ differ by ${\rm Supp}(\bfa;\bfb)=\{{j_1} , {j_2} , \ldots , j_{w( {\bfa}, {\bfb} )}\}$. 
Furthermore, for a set $\cU=\{u_1, u_2, \ldots, u_m \} \subseteq \{{1,\ldots,n\}}$, we let $(\bfa;\bfb)_{\cU} = ((a_{u_1}, a_{u_2}, \ldots, a_{u_m});(b_{u_1}, b_{u_2}, \ldots, b_{u_m})) \in \mathbb{F}_2^m \times \mathbb{F}_2^m,$ where $u_1 < u_2 < \cdots < u_m$.

Define $\cC_p(n) \subseteq \mathbb{F}_2^n \times \mathbb{F}_2^n$ according to a parity and a weight constraint
\begin{align*} 
\cC_p(n) := &\{ (\bfa;\bfb) \in \mathbb{F}_2^n \times \mathbb{F}_2^n : w(\bfa;\bfb) \geq 1, \sum_{i=1}^n a_i \equiv 0 \bmod 2 \}  \\
&\cup \, \{ (\bfa;\bfb) \in \mathbb{F}_2^n \times \mathbb{F}_2^n : w(\bfa;\bfb) =0   \}. 
\end{align*}

The subscript ``p'' in $\cC_p(n)$ serves to indicate that the underlying code enforces a simple parity-check constraint. As before, let $d_H(\bfx, \bfy)$ stand for the Hamming distance between two binary vectors $\bfx, \bfy \in \mathbb{F}_2^n$. Equipped with this notation, we may state the following result.
\vspace{-0.1in}
\begin{proposition}\label{prop:d2} Let the code $\cC_p(n)$ be defined as above, with $n \geq 1$. The code has $d_{\lambda=1} = 2$, and 
$$|\cC_p(n)|=2^{2n-1} + 2^{n-1}.$$
The code $\cC_p(n)$ has optimal code size for the given minimum ALD and code length.
\end{proposition}
\begin{IEEEproof} 
Consider an arbitrary code $\cC$ of length $n$ and with $d_{\lambda=1} = 2$. We partition the codewords in $\cC$ into subsets of the form 
\begin{align*}
&\cC_{T,\bfv}:=\{(\bfa;\bfb)\in \cC: {\rm Supp}(\bfa;\bfb)=T, \\
&(\bfa;\bfb)_{[n] \setminus T}=(\bfv;\bfv), \}
\end{align*}
where $T$ ranges over all subsets of $\{{1,\ldots,n\}}$ and where $\bfv \in \mathbb{F}_2^{n-|T|}$ is arbitrary. 
We observe that a word $(\bfa;\bfb)\in\cC_{T,\bfv}$ is completely determined by $\bfa$.
In particular, $b_j=a_j+1$, if $j\in T$, and $b_j=a_j$ otherwise.

If $T$ is empty, then $|\cC_{T,\bfv}|\le 2^n$ for all $\bfv$. 

We show next that for any non-empty $T$ and any $\bfv \in \mathbb{F}_2^{n-|T|}$, $|\cC_{T,\bfv}| \leq 2^{|T|-1}$. 

Let $k \in T$ be an arbitrarily chosen element. We proceed by considering the following set:
$$ \cC_{T,\bfv}' := \{ (\bfa;\bfb)_{T \setminus k} : (\bfa;\bfb) \in \cC_{T,\bfv} \}.$$
We claim that $|\cC_{T,\bfv}'| = |\cC_{T, \bfv}|$. Suppose that on the contrary, $|\cC_{T, \bfv}'| < |\cC_{T, \bfv}|$. Then, there exist two codewords $(\bfa;\bfb), (\bfc;\bfd) \in \cC_{T, \bfv}$ such that $(\bfa;\bfb)_{\{{1,\ldots,n\}} \setminus k} = (\bfc;\bfd)_{\{{1,\ldots,n\}} \setminus k}$. In other words, $(\bfa;\bfb)$ and $(\bfc;\bfd)$ only differ in one single position $k$. Furthermore, given that $k \in T$, it follows that $d_{\lambda=1}( (\bfa;\bfb), (\bfc;\bfd) ) = 1$, which contradicts the original assumptions as $(\bfa;\bfb), (\bfc;\bfd)$ cannot both belong to $\cC$. Hence, if $T$ is non-empty, $|\cC_{T, \bfv}| = |\cC_{T, \bfv}'| \le 2^{|T|-1},$ so that
\[|\cC|\le \sum_{i=1}^n \nchoosek{n}{i} \cdot 2^{n-i} \cdot 2^{i-1}+2^n= 2^{2n-1} + 2^{n-1}. \]

Note that $\cC_p(n)$ can detect a single Class 1 error, since a single Class 1 error changes the parity of $\bfa$ 
for any codeword in the first subset of $\cC_p(n)$.
Hence, $d_{\lambda=1}(\cC_p(n)) \geq 2$ and given that simple counting arguments show that $|\cC_p(n)|=2^{2n-1} + 2^{n-1}$, the result follows.
\end{IEEEproof}

The basic idea behind Proposition~\ref{prop:d2} was to use two sets, $\{ (\bfa;\bfb) \in \mathbb{F}_2^n \times \mathbb{F}_2^n : w(\bfa;\bfb) \geq 1 \}$ and $\{ (\bfa;\bfb) \in \mathbb{F}_2^n \times \mathbb{F}_2^n : w(\bfa;\bfb) = 0 \}$, the first of which imposes a parity constraint that increases the ALD. The second set was used to enlarge the resulting codebook. We next pursue a similar idea which relies on partitioning $\mathbb{F}_2^n \times \mathbb{F}_2^n$ into subsets that satisfy different constraints that boost either the ALD distance or the code size.

Our next immediate goal is to construct a family of codes $\cC(n)$ with $n = 2^v  - 2$, where $v \geq 5$, that have minimum ALD at least three, and size 
$$|\cC(n)| \geq \frac{4^n}{n+2} + 2^n.$$ 
Recall that a code with minimum ALD equal to three, for $\lambda=1$, can either (a) correct a single Class 1 error or (b) detect a single Class 2 error. 
We consider the following two sets of paired words $\cP_1$ and $\cP_2,$ where every $(\bfa;\bfb) \in \cP_1$ has the property that  
$w(\bfa;\bfb) \leq 7$ and $w(\bfa;\bfb) \equiv 1 \bmod 2$; and every $(\bfa;\bfb) \in \cP_2$ has the property that $w(\bfa;\bfb) \geq 9$. 

Let $\cC_H(i) \subseteq \mathbb{F}_2^i$ denote a binary code of length $i$ with minimum Hamming distance three. Recall the definition of the set $\mathcal{S}(\bfa; \bfb)$ from the introduction of this section and let
\begin{align*} \cC_1(n) :=& \Big \{ ({\bfa};{\bfb}) \in \cP_1 \text{ s.t. } \mathcal{S}(\bfa;\bfb) \in \cC_H(w({\bfa};{\bfb})) \Big \}, \text{ and} \end{align*}
\begin{align*} \cC_2(n) := \Big \{ ({\bfa};{\bfb}) \in \cP_2 \cap \cC_{l, \bfu}(n) \Big \},
\end{align*}
for $\cC_{l,\bfu}$ as defined in~\eqref{eq:cosetcode}, and for some $\bfu \in \mathbb{F}_2^v$. 

Furthermore, define $\cC(n) \subseteq \mathbb{F}_2^n \times \mathbb{F}_2^n$ as
\begin{align*} \cC(n) := \cC_1(n) \cup \cC_2(n).
\end{align*}

\begin{lemma} For any positive integer $v$, $d_{\lambda=1}(\cC(2^v-2)) \geq 3$. 
\end{lemma}
\begin{IEEEproof} Let $n = 2^v -2$. We show that $\cC(n)$ has minimum ALD three by demonstrating that 
$\cC(n)$ can either correct a single Class 1 error or detect a single Class 2 error.  

We start by establishing that $\cC(n)$ can correct a single Class 1 error. Suppose that $(\bfa;\bfb) \in \cC(n)$ 
was transmitted and that $(\bfc;\bfd) \in \mathbb{F}_2^n \times \mathbb{F}_2^n$ was received, where $(\bfc;\bfd)$ is the result of at most one Class 1 error occurring in $(\bfa;\bfb)$ at some position $j \in {\rm Supp}({\bfa};{\bfb})$. Note that  $d_H(\mathcal{S}(\bfa;\bfb),\mathcal{S}(\bfc;\bfd))\leq 1$. 
As a result, if $w(\bfc;\bfd) = i \leq 7$, one can use the decoder for $\cC_H(w({\bfc};{\bfd}))$ to correct the Class 1 error. Otherwise, if  $i \geq 9$, one can use the decoder for $\cC_{l, \bfu}(n)$ to correct the Class 1 error. 

Suppose that $(\bfa;\bfb) \in \cC(n)$ was transmitted and that $(\bfc;\bfd)$ was received, where $(\bfc;\bfd)$ is the result of at most one single Class 2 error occurring in $(\bfa;\bfb)$. Assume first that $w(\bfa;\bfb) = i \leq 7$. Since $w(\bfa;\bfb) \equiv 1 \bmod 2$, we have $w(\bfc;\bfd) \equiv 0 \bmod 2$ if a Class 2 error occurred, and $w(\bfc;\bfd) \equiv 1 \bmod 2$ otherwise. Thus, we can detect a Class 2 error whenever $w(\bfa;\bfb) = i \leq 7$. If $w(\bfa;\bfb) \geq 9$, the result follows from the arguments presented in Lemma~\ref{lem:construct1}.
\end{IEEEproof}

The next lemma provides a bound on the cardinality of $\cC(n)$.

\begin{lemma} For $v \geq 5$, $|\cC(2^v-2)| \geq \frac{4^n}{n+2} + 2^n$. \end{lemma}

\begin{IEEEproof} Using an averaging argument, one can prove that 
\begin{align}\label{eq:19card}
|\cC_1(n)| \geq 2^n \cdot \sum_{k\in\{1,3,5,7\}} \frac{ \nchoosek{n}{k} }{ 8 }.
\end{align}
More precisely, the claim follows by observing that a Hamming code of length $i$ and its coset codes partition the space $\mathbb{F}_2^i$; each of the $i+1$ coset codes has minimum distance three. Therefore, there exists binary codes $\cC^{(1)}_b, \cC^{(3)}_b, \cC^{(5)}_b, \cC^{(7)}_b$, of lengths $1,3,5,7$, respectively, each with minimum Hamming distance three, so that
$$|\cC^{(i)}| \geq \frac{\nchoosek{n}{i}}{i+1}, \, \text{ for } i \in \{1,3,5,7\}.$$ 
From $\cC^{(i)}_b$, $i \in \{1,3,5,7\}$, it is straightforward to derive a code $\cC_1(n) \subseteq \cP_1$ whose cardinality satisfies (\ref{eq:19card}).

Using a similar reasoning, there exists a $\bfu \in \mathbb{F}_2^v$ in the definition of $\cC_2(n)$ for which
\[ | \cC_2(n) | \geq 2^n \cdot \sum_{k=9}^n \frac{ \nchoosek{n}{k} }{n +2}, \;\; \text{so that}\]
\vspace{-0.1in}
\begin{align*}
| \cC(n)| &\ge 2^n \cdot \left( \sum_{k\in\{1,3,5,7\}} \frac{ \binom{n}{k} }{ 8} +  \sum_{k=9}^n \frac{ \binom{n}{k} }{n +2} \right)\\
&=\frac{2^n}{n+2}\cdot \left( \sum_{k=0}^n  \binom{n}{k} +\sum_{k\in\{1,3,5,7\}}\frac{(n+2)\binom{n}{k}}{8}\right) \\
&-\frac{2^n}{n+2}\cdot \left( \sum_{k=0}^8 \binom{n}{k}\right) \\
&= \frac{4^n}{n+2}+ \frac{2^n}{n+2}\,B(n),
\end{align*}
where
\[B(n) = \sum_{k\in\{1,3,5,7\}}\frac{(n+2)\binom{n}{k}}{8} - \sum_{k=0}^8 \binom{n}{k}.\]
Since $B(n) - (n+2) > 0$ for $n \geq 30$, the result follows.



\end{IEEEproof}

We consider next the case $d> 3$, with $\lambda =1$. The basic idea behind the construction is to map the symbols from $\mathbb{F}_2 \times \mathbb{F}_2$ to $\{0,1,2,3\}$ and then use codes in the Lee metric with proper modifications. 
The map of interest $\phi^{(4)}: \mathbb{F}_2 \times \mathbb{F}_2 \to \{0,1,2,3\}$ is defined as follows:

$$ \phi^{(4)}(0;0) \to 0,\; \phi^{(4)}(0;1) \to 1,\; \phi^{(4)}(1;0) \to 2,\; \phi^{(4)}(1;1) \to 3.$$

As before, for a word $(\bfa;\bfb) \in \mathbb{F}_2^n \times \mathbb{F}_2^n$, the map is extended in the natural manner according to $\phi^{(4)}(\bfa;\bfb)$ $= (\phi^{(4)}(a_1;b_1),$ $\ldots, \phi^{(4)}(a_n;b_n))$; the image of a set under the extended map $\phi^{(4)}$ is the set of images of elements of the set under the map $\phi^{(4)}$. 

Let $q$ be an odd prime and suppose that $\ell$ and $d\geq 2$ are positive integers. Let $u \in \mathbb{Z}_{d}$ and $\bfz \in \mathbb{F}_{q^{\ell}}^{\lfloor \frac{d}{2} \rfloor}$. Furthermore, suppose that $\alpha$ is a primitive element of $\mathbb{F}_{q^{\ell}}$. Then, for $n = q^{\ell} -1$, define $\cC_N(n,u,{\bfz})$ as follows:
\begin{align}
 \cC_N(n,u, \bfz) := \Big \{& ({\bfa};{\bfb}) \in \mathbb{F}_2^n \times \mathbb{F}_2^n : \label{eq:c1nduz}\\
 & \sum_{i=1}^n \phi^{(4)}(a_i ;b_i) \equiv u \mod d \nonumber \\
 & \sum_{i=1}^n \phi^{(4)}(a_i ;b_i) \cdot \alpha^{i} = z_1 \nonumber \\
 &\vdots \nonumber \\
 &\sum_{i=1}^n \phi^{(4)}(a_i ;b_i) \cdot \alpha^{i \cdot {\lfloor d/2 \rfloor}} = z_{\lfloor d/2 \rfloor} \}, \nonumber
\end{align}
with all operations performed over $\mathbb{F}_{q^{\ell}}$. 

\begin{lemma}\label{lem:constr} Let $q$ be an odd prime where $q \geq d+1$. For $n=q^{\ell} -1$, $d$ an odd positive integer, $u \in \mathbb{Z}_d$ and $\bfz \in \mathbb{F}_{q^\ell}^{\lfloor \frac{d}{2} \rfloor}$, $d_{\lambda=1}(\cC_N(n,u,{\bfz})) \geq d$. 
\end{lemma}
\begin{IEEEproof} Suppose that $({\bfa};{\bfb}), ({\bfc};{\bfd}) \in \cC_{N}(n,u, {\bfz})$. For simplicity, let $\overline{{\rm Supp}}(\bfa;\bfb)$ stand for $[n] \setminus {\rm Supp}(\bfa;\bfb)$. Furthermore, let
\begin{align*}
I &= | \{ m \in {\rm Supp}(\bfa;\bfb) \cap {\rm Supp}(\bfc;\bfd) : a_m \neq c_m \}|, \\
J &= | \{ m \in \overline{{\rm Supp}}(\bfa;\bfb) \cap \overline{{\rm Supp}}(\bfc;\bfd) : a_m\neq c_m \}|, \text{ and} \\
K &= | \{ m \in [n]: (a_m, b_m) \neq (c_m, d_m)  \} | - (I+J).
\end{align*}
In order to show that $d_{\lambda=1}(({\bfa};{\bfb}), ({\bfc};{\bfd})) \geq d$, we need to prove that $I + 2 K + 4 J \geq d$. 
Clearly, if over $\mathbb{F}_q$ the Lee distance of $\phi^{(4)}(\cC_N(n,u, {\bfz}))$ is at least $d$, then $$I + 2K + 4J \geq I + 2K + 3J \geq d$$ and the result holds.

To see that the Lee distance of $\cC_N(n,u, {\bfz})$ is at least $d$, notice that we can recover any error vector of weight at most $\lfloor \frac{d}{2} \rfloor$ from the power sums listed in (\ref{eq:c1nduz}). More precisely, if the Lee weight of $\phi^{(4)}({\bfe};{\bff})$ is at most $\lfloor \frac{d}{2} \rfloor$, where $({\bfe};{\bff})$ denotes the error vector, it is known from~\cite{R06} that given $\sum_{i=1}^n \phi^{(4)}(e_i ;f_i) \cdot \alpha^{i \cdot k}$ for $k=\{0,1, \ldots, \lfloor \frac{d}{2} \rfloor \}$, one can uniquely determine the vector $\phi^{(4)}(\bfe;\bff)$. Clearly, from (\ref{eq:c1nduz}), we have the values of $\sum_{i=1}^n \phi^{(4)}(e_i; f_i) \cdot \alpha^{i \cdot k}$ for $k \in \{ 1, \ldots, \lfloor \frac{d}{2} \rfloor \}$. For $k=0$, we can uniquely determine $\sum_{i=1}^n \phi^{(4)}(e_i; f_i)$ as by assumption, the Lee weight of $\phi^{(4)}(\bfe;\bff)$ is at most 
$\lfloor \frac{d}{2} \rfloor$.
\end{IEEEproof}

Using an averaging argument that exploits the fact that the code from (\ref{eq:c1nduz}) partitions the space into 
$d(n+1)^{\lfloor d/2 \rfloor}$ sets of codewords, we arrive at
\begin{align}\label{eq:codesize20}
| \cC_N(n,u,{\bfz}) | \geq \frac{4^n}{d (n+1)^{\lfloor d/2 \rfloor} },
\end{align}
which for $d \geq 7$ improves the lower bounds from the previous section. {Recall from Section~\ref{sec:constructions} that one alternative to using codes in the ALD is to construct codes in the Hamming distance over $\mathbb{F}_4$. For instance when $d=3$, a code with minimum Hamming distance $3$ can correct any error with Hamming weight one and so it has codewords which are at distance at least $3$ under the ALD. Since the largest code of distance $d = 2r+1$ in the Hamming metric over $\mathbb{F}_4$ has cardinality $\frac{4^n}{\sum_{j=0}^r 3^j \cdot \nchoosek{n}{j}}$ we see that for $d \geq 9$, and $n$ large enough the quantity in (\ref{eq:codesize20}) is larger than the largest possible code with Hamming distance $d$ over $\mathbb{F}_4$.} From Proposition~\ref{prop:simple}, we also conclude that
\[ \frac{4^n}{d(n+1)^{r}} \leq A_\lambda(n,2r+1)\le 2^n \cdot \sum_{\ell=0}^n \frac{\binom{n}{\ell}}{ \sum_{j=0}^{r} \nchoosek{\ell}{j}},\]
where $d = 2r+1$. Note that the gap between the upper and lower bounds is a function of $r$ as well as $n$.
Table V compares the lower bounds of (\ref{eq:c1nduz}) with the LP bound of (\ref{eq:constraint}). 

The last construction we present may be used to generate codes under the ALD for a large range of parameters. In particular, although our focus is on codes with integer $\lambda$ values, we observe that the construction in question may also be used to generate ALD codes with $0 < \lambda < 1$.


\begin{table}
\centerline{
\begin{tabular}{c|c c c }
$n\backslash d_{\lambda=1}$ & 3 & 5 & 7 \\
\hline
1 & 1--3  \\
2 & 2--9 & 1--2 & 1--2  \\
3 & 6--30 & 1--7 & 1--5  \\
4 & 18--99 & 3--21 & 1--13 \\
5 & 57--336 & 6--67 & 1--35 \\
6 & 196--1161 & 17--219 & 2--101  \\
7 & 683--4080 & 52--731 & 5--296  \\
8 & 2428--14535 & 162--2483 & 13--895  \\
9 & 8739--52377 & 525--8563 & 38--2783  \\
10 & 31776--190557 & 1734--29901 & 113--8890 \\
\end{tabular}}
{\caption{Comparison of the upper bound based on the LP in (\ref{eq:constraint}) (value on the right of the dash line) with (\ref{eq:codesize20}) (value on the left of the dash line). The relative gap increases with increasing ALD.}}
\end{table}

Let $\cC_{M}(n,d)$ be a code over $\mathbb{F}_3$ of length $n$ with minimum Manhattan ($\ell_1$) distance $d$. Similarly, let 
$\cC_H(n,d)$ denote a binary code of length $n$ and minimum Hamming distance $d$. Define
\begin{align}
\cC_{\lambda}(n,d) :=& \{ ( {\bfa};{\bfb} ) \in \{0,1\}^n \times \{0,1\}^n : \nonumber \\
& (a_1+b_1, a_2 + b_2, \ldots, a_n + b_n) \in \cC_{M}(n,\lceil \frac{d}{1+\lambda} \rceil), \notag \\
& \mathcal{S}({\bfa};{\bfb}) \in \cC_H({w({\bfa};{\bfb})},\lceil \frac{d}{\lambda} \rceil) \}, \label{eq:e2}
\end{align}
where addition is performed over the reals.
We have the following lemma.

\begin{lemma} Given integers $n$ and $d$, one has $d_{\lambda}(\cC_{\lambda}(n,d)) \geq d$. \end{lemma}
\begin{IEEEproof} Suppose that $({\bfa};{\bfb}), ({\bfc};{\bfd}) \in \cC_{\lambda}(n,d)$, and let 
\begin{align*}
I &= | \{ m \in {\rm Supp}(\bfa;\bfb) \cap {\rm Supp}(\bfc;\bfd) : a_m \neq c_m \}|,\\
J &= | \{ m \in \overline{{\rm Supp}}(\bfa;\bfb) \cap \overline{{\rm Supp}}(\bfc;\bfd) : a_m\neq c_m \}|, \text{ and} \\
K &=| \{ m \in [n] : (a_m, b_m) \neq (c_m, d_m)  \} | - (I+J).
\end{align*}
Recall that if $d_{\lambda}(({\bfa};{\bfb}), ({\bfc};{\bfd})) \geq d$, we need to show that 
$$\lambda \cdot I + (1 + \lambda)\cdot K + 2(1+\lambda)\cdot J \geq d.$$ 
If $(a_1+b_1, \ldots, a_m + b_m) \neq (c_1+d_1, \ldots, c_m + d_m)$, then $K+2J \geq \lceil \frac{d}{1+\lambda} \rceil$ from (\ref{eq:e2}). Thus, 
$$ {d \leq} (1+\lambda)\cdot (K+2J) {=} (1 + \lambda)\cdot K + 2(1+\lambda)\cdot J,$$
which proves the claimed result. Otherwise, if $(a_1+b_1, \ldots, a_m + b_m) = (c_1+d_1, \ldots, c_m + d_m)$, the result is an immediate consequence of (\ref{eq:e2}).
\end{IEEEproof}

\section{Conclusion}
We introduced a novel distance measure suitable for modeling the confusion graph of a DNA-based data storage readout channel. The distance,
termed the asymmetric Lee distance (ALD) is related to the Lee distance, but refines the Lee distance in terms of allowing symbol weightings and  asymmetries in the symbol distances. We derived upper bounds on the largest size of ALD codes using LP-based techniques, and 
provided linear and non-linear code constructions for some choices of symbol weighting parameters.

\vspace{-0.1in}

\end{document}